\documentclass[twocolumn,english,footinbib,bibnotes,superscriptaddress]{revtex4-1}
\usepackage{graphicx}
\usepackage[usenames,dvipsnames]{color}
\usepackage{dcolumn}
\usepackage{bbm}
\usepackage{bm}
\usepackage{amsmath}
\usepackage{amssymb}
\usepackage{times}
\usepackage{placeins}
\usepackage{hyperref}
\hypersetup{colorlinks=false}
\usepackage[version=3]{mhchem}
\usepackage{csquotes}

\newcommand{\RNum}[1]{\uppercase\expandafter{\romannumeral #1\relax}}

\newcommand{\eqs}{\hspace{0.5cm}}

\newcommand{\nodagger}{{\phantom\dagger}}
\newcommand{\D}[1]{\frac{\mathrm{d}}{\mathrm{d}{#1}}}

\graphicspath{{figures/compressed/}}

\begin{document}

\title{Functional renormalization group for frustrated magnets with nondiagonal spin interactions}
\author{Finn Lasse Buessen}
\affiliation{Institute for Theoretical Physics, University of Cologne, 50937 Cologne, Germany}
\author{Vincent Noculak}
\affiliation{Dahlem Center for Complex Quantum Systems and Fachbereich Physik, Freie Universit\"at Berlin, 14195 Berlin, Germany}
\author{Simon Trebst}
\affiliation{Institute for Theoretical Physics, University of Cologne, 50937 Cologne, Germany}
\author{Johannes Reuther}
\affiliation{Dahlem Center for Complex Quantum Systems and Fachbereich Physik, Freie Universit\"at Berlin, 14195 Berlin, Germany}
\affiliation{Helmholtz-Zentrum f\"ur Materialien und Energie, Hahn-Meitner-Platz 1, 14019 Berlin, Germany}
\date{\today}

\begin{abstract}
In the field of quantum magnetism, the advent of numerous spin-orbit assisted Mott insulating compounds, such as the family of Kitaev materials, has led to a growing interest in studying general spin models with non-diagonal interactions that do not retain the SU(2) invariance of the underlying spin degrees of freedom. However, the exchange frustration arising from these non-diagonal and often bond-directional  interactions for two- and three-dimensional lattice geometries poses a serious challenge for numerical many-body simulation techniques. 
In this paper, we present an extended formulation of the pseudo-fermion functional renormalization group that is capable of capturing the physics of frustrated quantum magnets with generic (diagonal and off-diagonal) two-spin interaction terms. 
Based on a careful symmetry analysis of the underlying flow equations, we reveal that the computational complexity grows only moderately, as compared to models with only diagonal interaction terms. 
We apply the formalism to a kagome antiferromagnet which is augmented by general in-plane and out-of-plane Dzyaloshinskii-Moriya (DM) interactions, as argued to be present in the spin liquid candidate material herbertsmithite. We calculate the complete ground state phase diagram in the strength of in-plane and out-of-plane DM couplings, and discuss the extended stability of the spin liquid of the unperturbed kagome antiferromagnet in the presence of these couplings.
\end{abstract}

\maketitle


\section{Introduction}
\label{sec:intro}

The study of quantum magnets has produced an impressive streak of deep conceptual insights, often with implications that go far beyond the scope of the field of magnetism. An early revelation was Haldane's conjecture \cite{Haldane1983a,Haldane1983b} of the fundamental difference between integer and half-integer spin chains, whose gapped/gapless energy spectra he explained using topological terms -- laying the conceptual groundwork for what has later been coined symmetry-protected topological states of matter \cite{SPT}. Another striking example is the generalization of the Lieb-Schultz-Mattis theorem \cite{LSM} to higher dimensions by Oshikawa \cite{Oshikawa2000} and Hastings \cite{Hastings2004}, stating in particular that two-dimensional spin-1/2 systems cannot be featureless (for an odd number of spins per unit cell), 
i.e.~if the energy spectrum is gapped, then the system must exhibit topological order and a non-trivial ground-state degeneracy. 
This general signature of intrinsic topological order is not limited to magnetic systems, but in fact the general consequence of the formation of long-range entanglement \cite{Chen2010}. Signatures for such macroscopic entanglement  \cite{LevinWen2006,KitaevPreskill2006} also allow, for instance, to positively identify quantum spin liquids \cite{Savary2016} -- long elusive ground states of quantum magnets that defy any classical ordering tendencies, but instead exhibit quantum order \cite{Wen2002}, concurrent with a fractionalization of the original spin degrees of freedom and the emergence of a lattice gauge structure.

A common thread in this foundational work on quantum magnets is that their microscopic spin interactions are typically written in terms of Heisenberg models that retain the full SU(2) invariance of the underlying spins. The reason to do so 
can arguably be traced back to what has been one of the core motivations to study quantum magnetism in the first place: the discovery of high-temperature superconductivity  \cite{Bednorz1986}
in close proximity to a ``parent" Mott insulating state in which the charge degrees of freedom are frozen out and the remaining local moments are spin degrees of freedom -- a quantum magnet. To derive the microscopic spin exchange in such Mott insulators one typically expands the original electronic Hubbard model in terms of the (considerably suppressed) electronic hopping, which directly leads to the aforementioned SU(2)-invariant Heisenberg model \cite{Auerbach1998}.
In recent years, however, there has been a flurry of activity directed towards the analysis of quantum magnets with interactions that explicitly break the SU(2) spin rotational symmetry. The most prominent example is the Kitaev model \cite{Kitaev2006}, in which SU(2) spin-1/2 degrees of freedom interact via bond-directional exchanges, i.e.~Ising-like interactions where the easy axis of the magnetic exchange depends on the spatial orientation of the exchange bond. Kitaev's exact analytic derivation of a number of quantum spin liquid ground states for this model \cite{Kitaev2006} has led to an intense search for ``Kitaev materials" \cite{Trebst2017} that give rise to such bond-directional exchanges. Guided by the work of Khaliullin, Jackeli, and coworkers \cite{Jackeli2009}, a number of spin-orbit assisted Mott insulators \cite{WitczakKrempa2014} have been explored as candidate materials, including 
Na$_{\rm 2}$IrO$_{\rm 3}$,  
($\alpha,\beta,\gamma$)-Li$_{\rm 2}$IrO$_{\rm 3}$, 
and RuCl$_{\rm 3}$ amongst others \cite{Rau2016}.
What is, however, common to all these materials is that their microscopic description includes not only symmetric (Kitaev- and Heisenberg-like) exchange interactions but also off-diagonal, bond-directional exchanges such as the so-called $\Gamma$ terms, 
i.e.~their generic spin models are often written as
\begin{equation}
	H = -\sum_{\gamma \rm-bonds} J \,\, {\bf S}_i {\bf S}_j + K \,\, S_i^{\gamma} S_j^{\gamma} 
			+ \Gamma \left(  S_i^{\alpha} S_j^{\beta} + S_i^{\beta} S_j^{\alpha} \right) \,,
	\label{eq:HKG-model}
\end{equation}
where the precise coupling strengths, of course, depend on the actual compound at hand (see Ref.~\onlinecite{Winter2016} for an overview of Kitaev materials) and which are sometimes augmented by further terms, such as another form of off-diagonal $\Gamma'$ interactions \cite{Winter2016} or a Dzyaloshinskii-Moriya exchange.  

It is the purpose of this manuscript, to expand one of the few numerical many-body approaches capable of studying frustrated quantum magnets in two or three spatial dimensions, the pseudo-fermion functional renormalization group (pf-FRG) \cite{Reuther2010}, such that it can efficiently treat Hamiltonians such as the one in Eq.~\eqref{eq:HKG-model}, or more generally, arbitrary two-spin interactions of the form
\begin{equation}
	H=\sum\limits_{ij}  {\bf S}_i 
\begin{pmatrix} K_{xx} & \Gamma_{xy} & \Gamma_{xz}  \\  \Gamma_{yx}  & K_{yy} & \Gamma_{yz} \\ \Gamma_{zx}& \Gamma_{zy}& K_{zz}\\  \end{pmatrix}	
	{\bf S}_j \,.
	\label{eq:pffrg:hamiltonian}
\end{equation}
This has remained a major technical challenge thus far, as the approach was initially derived for SU(2) invariant Heisenberg models \cite{Reuther2010} by decomposing the original spin degrees of freedom into auxiliary Abrikosov fermions (or pseudo-fermions) and then employing the well-known fermionic functional renormalization group (FRG) approach introduced by Wetterich \cite{Wetterich1993}. Going beyond the Heisenberg interactions, by including Kitaev-type interactions \cite{Reuther2011c} or a Dzyaloshinskii-Moriya exchange \cite{Hering2017}, amounted to the tedious exercise of rederiving, for each coupling type, the fermionic flow equations at the heart of the FRG approach. Including other types of interactions such as the off-diagonal $\Gamma$-exchange has also been hindered by the expectation that the number of flow equations that need to be handled numerically grows tremendously (by some three orders of magnitude) for such more general interaction types.  

In what follows, we will overcome this challenge and provide an efficient parametrization for the pseudo-fermion functional renormalization group, which is suited to numerically investigate general spin exchanges of the above form with only a moderate increase in computational cost. This progress is based on a careful symmetry analysis of the flow equations derived for these more general interaction types. 
In particular, we argue that in the presence of time-reversal symmetry, the increase in computational complexity remains moderate, and the overall complexity is two orders of magnitude smaller than in the general case that allows breaking of time-reversal symmetry.

We corroborate the expanded usability of the method by an exemplary study of the Heisenberg antiferromagnet on the kagome lattice with additional in-plane and out-of-plane Dzyaloshinskii-Moriya (DM) interactions -- off-diagonal spin interactions that in prior implementations of the pf-FRG algorithm could only be handled for specialized cases and using significant numerical resources
\cite{Hering2017}. 
With the symmetry-constrained pf-FRG implementation introduced here, we demonstrate the increased numerical efficiency
by mapping out an entire phase diagram (in the in-plane and out-of-plane DM coupling strengths).
Physically, such DM interactions have been argued \cite{Zorko2008,Rigol2007} to be present, for instance, in the spin liquid candidate material herbertsmithite \cite{Norman2016}. 
We find that the spin liquid ground state of the unperturbed kagome Heisenberg antiferromagnet, indicated in pf-FRG calculations \cite{Suttner2014,Buessen2016}, 
is robust against small out-of-plane DM interactions up to $D/J \approx 0.1$. We further find that additional in-plane DM interactions only have a comparably small impact on the phase diagram -- the model exhibits an extended spin liquid regime for realistic parameter estimates that go up to $D'/J\approx 0.3$ in herbertsmithite. 

The remainder of the manuscript is structured as follows. 
In Section \ref{Sec:TRinvariant}, we present a concise overview of the pseudo-fermion functional renormalization group (pf-FRG) approach to general, time-reversal invariant quantum spin models, summarizing our most important results for the practical application of the method. 
In the remaining sections, based on a careful symmetry analysis, we present the detailed derivation of the pf-FRG approach.
We point out some fundamental differences between the FRG approach applied to (symmetry-constrained) pseudo-fermions and (conventional) fermions, which we summarize in a systematic classification of the projective symmetries of the pseudo-fermions in Section \ref{sec:symmetries}. 
This, in turn, allows to provide an efficient, symmetry-constrained vertex parametrization in Section \ref{sec:VertexParametrization} and discuss the general symmetries of the pf-FRG flow equations in Section \ref{sec:symmetriesFlowEquation}. 
An application example to the kagome antiferromagnet with additional in- and out-of-plane DM interactions is presented in Section \ref{sec:application}, followed by conclusions and a brief outlook in Section \ref{sec:conclusions}.


\begin{figure*}[t]
\includegraphics[width=0.95\linewidth]{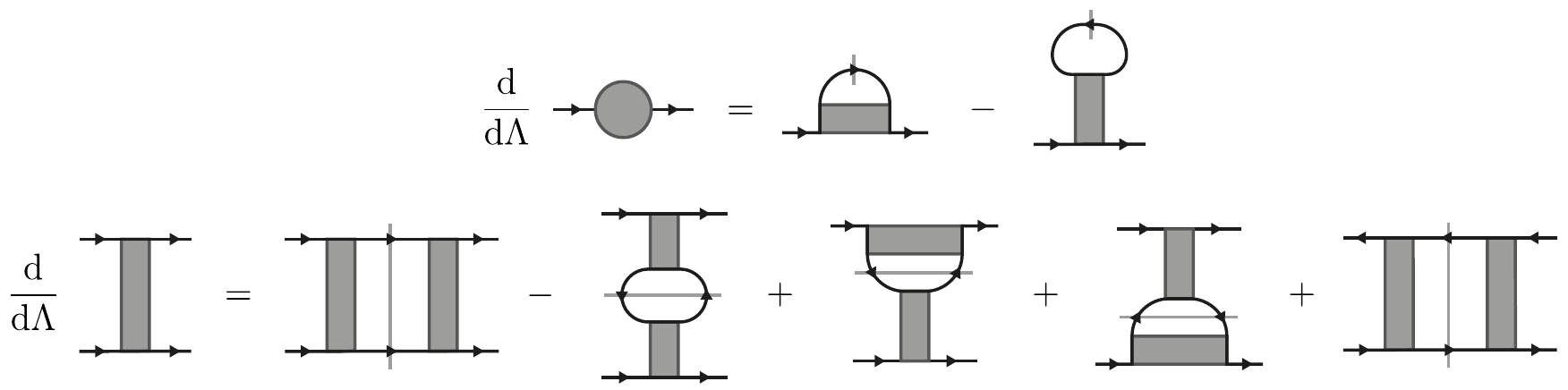}
\caption{{\bf Flow equations} for the one-particle (top row) and two-particle vertices (bottom row). The diagram for the one-particle vertex should be read as $\Sigma(\omega) \delta_{\alpha'\alpha}\delta_{i'i}\delta_{\omega'\omega}$ and the diagram for the two-particle vertex represents the expression $\Gamma^{\mu\nu}_{i_1i_2}(s,t,u) \sigma^\mu_{\alpha_{1'}\alpha_{1}} \sigma^\nu_{\alpha_{2'}\alpha_{2}} \delta_{\omega_{1'}+\omega_{2'}-\omega_{1}-\omega_{2}}$ (see text for details). In the single-particle flow equation, the slashed propagator line represents the single-scale propagator. In the two-particle flow equation, the pair of slashed propagator lines represents the two terms $G(\omega_1)S_\mathrm{kat}(\omega_2)+G(\omega_2)S_\mathrm{kat}(\omega_1)$. The lattice site index is preserved along the solid black lines, such that only the second term on the right hand side of the top row and the second term in the bottom row contain internal summations over lattice sites. A more explicit, yet significantly lengthier, representation of the flow equations is given in Ref.~\cite{Buessen2019a}}
\label{fig:pffrg:flowequation}
\end{figure*}

\section{Functional renormalization group for time-reversal invariant systems: Overview}
\label{Sec:TRinvariant}

In general terms, the pf-FRG approach \cite{Reuther2010} is a two-step scheme of (i) re-writing SU($N$) spin operators in terms of auxiliary Abrikosov fermions (pseudo-fermions), followed by (ii) the application of a fermionic FRG scheme \cite{Wetterich1993}. This overall pf-FRG formalism is by now well established and has been studied extensively in the past, establishing for instance that the approach becomes exact in the independent limits of large $S$ \cite{Baez2017} and large $N$ \cite{Buessen2018a,Roscher2018} (on a mean-field level) and demonstrating its general applicability also to three-dimensional frustrated quantum magnets, where results have been seen to be in good agreement with predictions from quantum Monte Carlo calculations \cite{Iqbal2016} or series expansions \cite{Buessen2016}.
Here we demonstrate, in a careful symmetry analysis, that the combination of a fermionic FRG approach with a pseudo-fermionic model (instead of a regular fermion model) leads to a considerable simplification of the underlying renormalization group flow equations, owed to an extended set of projective symmetries \cite{Wen2002} that is present in every pseudo-fermionic Hamiltonian per construction. 

We put our focus on time-reversal invariant spin models, which further augments the symmetry constraints, and consider microscopic models comprising arbitrary off-diagonal two-spin interactions, such as those given in Eq.~\eqref{eq:pffrg:hamiltonian} above. Our discussion will not be limited to a particular spatial dimension -- in particular, our implementation of a symmetry-constrained pf-FRG approach (to be discussed below) will be capable to study, in a straight-forward manner, three-dimensional frustrated quantum magnets, for which efficient numerical many-body approaches are particularly scarce.

In a first step, we re-cast the spin operators in terms of pseudo-fermion operators
\begin{equation}
\label{eq:pffrg:pseudo-fermions}
S_i^\mu \rightarrow \frac{1}{2}f^\dagger_{i\alpha} \sigma_{\alpha\beta}^\mu f^\nodagger_{i\beta} \eqs,
\end{equation}
which is a faithful mapping under the half-filling constraint 
\begin{equation}
	\label{eq:half-filling}
	f^\dagger_{i\alpha} f^\nodagger_{i\alpha} = 1\eqs.
\end{equation} 
In the context of pseudo-fermionized spin models, it is a good approximation to fulfill this constraint only {\em on average} by setting the chemical potential to zero \cite{Reuther2010}. 
Yet, in principle, the constraint can also be fulfilled exactly by introducing an artificial imaginary chemical potential according to the Popov Fedotov scheme \cite{Popov1988,Roscher2019}.
Performing the substitution scheme \eqref{eq:pffrg:pseudo-fermions} on the spin Hamiltonian \eqref{eq:pffrg:hamiltonian}, one obtains a purely quartic Hamiltonian acting on the pseudo-fermion Hilbert space. 
The absence of any quadratic term makes the Hamiltonian inaccessible to perturbative approaches around a Gaussian theory. Instead, the functional renormalization group scheme is applied as a means to concurrently treat millions of flowing parameters in a renormalization group scheme. 
For pseudo-fermionic models, one can reason that this approach amounts to a simultaneous expansion in spin length (i.e.~a large-$S$ expansion) and in the spins' symmetry group SU($N$) (i.e.~a large-$N$ expansion).
Most importantly, despite being a formally uncontrolled approximation, the approach becomes separately exact, on the level of mean field considerations, in the two limits of large $S$ \cite{Baez2017} and large $N$ \cite{Buessen2018a,Roscher2018}. 
These two cases mark the classical limit, which is a suitable description for magnetically ordered states, and the limit of artificially enhanced quantum fluctuations, which is a good picture to capture spin liquid states. 
It is generally believed that the incorporation of these two channels on equal footing contributes to the pf-FRG's success in modeling the competition between magnetic order and spin liquid behavior in a number of model applications. 

The flow equations at the heart of the pf-FRG approach are obtained as a special case of the well known general fermionic FRG equations \cite{Wetterich1993}. 
Given the field-theoretical action of a (pseudo-)fermionic model, they are most conveniently formulated in terms of the one-line irreducible interaction vertices. 
Neglecting three-particle vertices and higher, the flow equations for the single-particle $\Sigma$  and the two-particle vertex $\Gamma$ are given by 
\begin{equation}
\label{eq:pffrg:flow:1particle}
\frac{d}{d\Lambda}\Sigma^\Lambda(1';1)=-\frac{1}{2\pi}\sum\limits_2 \Gamma^\Lambda(1',2;1,2)S^\Lambda(\omega_2)
\end{equation}
and
\begin{align}
\label{eq:pffrg:flow:2particle}
&\frac{d}{d\Lambda}\Gamma^\Lambda(1',2';1,2) \nonumber\\
&=\frac{1}{2\pi}\sum\limits_{3,4} \Big[ \Gamma^\Lambda(1',2';3,4)\Gamma^\Lambda(3,4;1,2) \nonumber\\
&\quad-\Gamma^\Lambda(1',4;1,3)\Gamma^\Lambda(3,2';4,2) - (3\leftrightarrow 4) \nonumber\\
&\quad+\Gamma^\Lambda(2',4;1,3)\Gamma^\Lambda(3,1';4,2) + (3\leftrightarrow 4) \Big] \nonumber\\
&\quad \times G^\Lambda(\omega_3)S_\mathrm{kat}^\Lambda(\omega_4) \eqs, 
\end{align}
respectively. Here the numbers $n=\{i_n, w_n, \alpha_n \}$ represent tuples of a lattice site index $i_n$, the Matsubara frequency $\omega_n$, and a spin index $\alpha_n$, respectively. 
Note that in formulating these flow equations, we have already used that the single-scale propagator $S$ and the full propagator $G$ are diagonal in all their arguments and  depend only  on the frequency argument. 
The renormalization group flow is then generated by a sharp cutoff function in frequency space, such that the full propagator is given by 
\begin{equation}
G^\Lambda(\omega)=\frac{\theta(|\omega|-\Lambda)}{i\omega-\Sigma^\Lambda(\omega)}
\end{equation}
and the single-scale propagator is given by 
\begin{equation}
S^\Lambda(\omega)=\frac{\delta(|\omega|-\Lambda)}{i\omega-\Sigma^\Lambda(\omega)} \eqs.
\end{equation}
In the flow equtions for the two-particle vertex, the single scale propagator is treated according to the Katanin truncation scheme \cite{Katanin2004}, 
\begin{equation}
S_\mathrm{kat}^\Lambda(\omega)=S^\Lambda(\omega) - \left( G^\Lambda(\omega) \right)^2 \D{\Lambda}\Sigma^\Lambda(\omega) \eqs.
\end{equation}
Note that the diagonal structure of the propagators in their spin arguments is inherited from the diagonal structure of the self-energy $\Sigma$, which in turn is a consequence of time-reversal symmetry in the pseudo-fermion Hamiltonian. 
By a more detailed symmetry analysis (Sec.~\ref{sec:symmetries}), we will see that the self-energy can be efficiently parametrized as 
\begin{equation}
\label{eq:pffrg:parametrization2point}
\Sigma(i'\omega'\alpha' ; i\omega\alpha) = \Sigma(\omega) \delta_{\alpha'\alpha}\delta_{i'i}\delta_{\omega'\omega} \eqs,
\end{equation}
where the basis function $\Sigma(\omega)$ obeys the symmetry relations
\begin{align}
\label{eq:pffrg:parametrization2point:1}
\Sigma(\omega) &\in i\mathbb{R} \nonumber\\
\Sigma(\omega) &= -\Sigma(-\omega) \eqs.
\end{align}
The two-particle vertex is parametrized as 
\begin{align}
\label{eq:pffrg:parametrization4point}
&\Gamma(1', 2'; 1, 2) = \nonumber\\
&\quad\left[ \left( \Gamma^{\mu\nu}_{i_1i_2}(s,t,u) \sigma^\mu_{\alpha_{1'}\alpha_{1}} \sigma^\nu_{\alpha_{2'}\alpha_{2}} \right) \delta_{i_{1'} i_1}\delta_{i_{2'} i_2} - (1' \leftrightarrow 2' ) \right] \nonumber\\
&\quad\times\delta_{\omega_{1'}+\omega_{2'}-\omega_{1}-\omega_{2}} \eqs, 
\end{align}
where $\sigma^0$ is the identity matrix and $\sigma^1$, $\sigma^2$, and $\sigma^3$ denote the usual spin Pauli matrices. Furthermore, we have introduced the transfer frequencies 
\begin{align}
s &= \omega_{1'}+\omega_{2'} \nonumber\\
t &= \omega_{1'}-\omega_{1} \nonumber\\
u &= \omega_{1'}-\omega_{2} \eqs. 
\end{align}
As the key property leading to a significant increase in numerical efficiency, the basis functions $\Gamma^{\mu\nu}_{i_1i_2}(s,t,u)$ obey the symmetry relations
\begin{align}
\label{eq:pffrg:parametrization4point:1}
\Gamma^{\mu\nu}_{i_1i_2}(s,t,u) &\in 
\left\{ \begin{array}{l}
\phantom i\mathbb{R} \quad \mbox{if $\xi(\mu)\xi(\nu)=1$} \\
i\mathbb{R} \quad \mbox{if $\xi(\mu)\xi(\nu)=-1$} 
\end{array} \right. \nonumber\\
\Gamma^{\mu\nu}_{i_1i_2}(s,t,u)&= \Gamma^{\nu\mu}_{i_2i_1}(-s,t,u) \nonumber\\
\Gamma^{\mu\nu}_{i_1i_2}(s,t,u)&= \xi(\mu)\xi(\nu)\Gamma^{\mu\nu}_{i_1i_2}(s,-t,u) \nonumber\\
\Gamma^{\mu\nu}_{i_1i_2}(s,t,u)&= \xi(\mu)\xi(\nu) \Gamma^{\nu\mu}_{i_2i_1}(s,t,-u) \nonumber\\
\Gamma^{\mu\nu}_{i_1i_2}(s,t,u)&= -\xi(\nu) \Gamma^{\mu\nu}_{i_1i_2}(u,t,s) \eqs, 
\end{align}
where 
\begin{equation}
\xi(\mu)= \left\{ \begin{array}{rl}
+1 &\mbox{if $\mu=0$} \\
-1 &\mbox{otherwise} \end{array} \right. \eqs.
\end{equation}

The initial conditions for the flow equations (\ref{eq:pffrg:flow:1particle},\ref{eq:pffrg:flow:2particle}) in the limit of large cutoff $\Lambda\to\infty$ are given by 
\begin{align}
\Sigma^{\Lambda\to\infty}(\omega)&=0 \nonumber\\
\Gamma_{ij}^{\Lambda\to\infty,\mu\nu}(s,t,u)&=\frac{1}{4}J^{\mu\nu}_{ij} \eqs,
\end{align}
where $J_{ij}^{\mu\nu}$ is the coupling constant defined by the general form of a two-spin Hamiltonian 
$H=\sum_{ij} J_{ij}^{\mu\nu} S_i^\mu S_j^\nu$.
Employing this parametrization, one obtains flow equations for the basis functions $\Sigma(\omega)$ and $\Gamma^{\mu\nu}_{i_1i_2}(s,t,u)$. The equations are diagrammatically shown in Fig.~\ref{fig:pffrg:flowequation}.
Note that the flow equations are formally derived at zero temperature, but it has been demonstrated \cite{Iqbal2016} that from such a zero temperature solution one can nevertheless extract finite temperature properties by relating the frequency cutoff $\Lambda$ to the actual temperature via $T=\frac{\pi}{2}\Lambda$.  

Once the flow equations have been solved numerically, and all vertex values are known, one may extract observables like the (static) spin-spin correlations $\chi^{\mu\nu}_{ij}=\langle S_i^\mu S_j^\nu \rangle$, which diagrammatically is given by 
\begin{equation}
\label{eq:pffrg:measurement}
\includegraphics[width=0.95\linewidth]{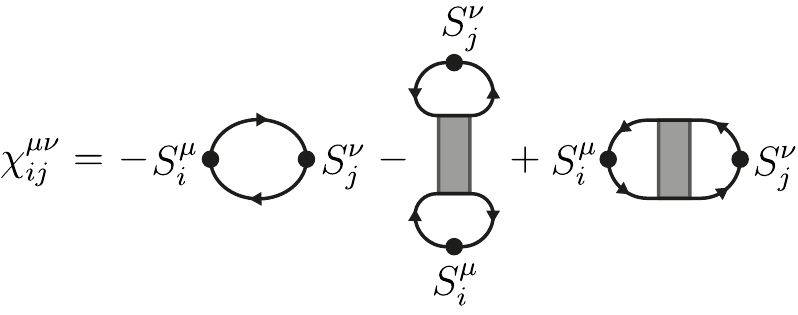} .
\end{equation}
A phase transition, accompanied by the spontaneous breaking of symmetries, is formally detected via a divergence in the RG flow of the vertex functions -- in the case of magnetic long-range order implying also a divergence in the corresponding spin correlations defined in Eq.~\eqref{eq:pffrg:measurement}. 
In practice, resulting from the truncation of the flow equations and finite numerical resolution, the divergence is often regularized to manifest only as a kink or a cusp in the RG flow (c.f. Fig.~\ref{fig:application:flowComparison}). 
Different schemes to improve the resolution of the phase transition are subject of current research \cite{Hering2017,Keles2018,Kiese2019}, addressing also the detection of valence bond solid configurations \cite{Reuther2010}.

It is worth mentioning that in the case of more symmetric spin models (e.g. Heisenberg interactions or Kitaev interactions), the parametrization reduces to previously known cases. 
For Heisenberg models, only the $\Gamma^{00}$ and $\Gamma^{11}=\Gamma^{22}=\Gamma^{33}$ components are non-zero, and other terms cannot be generated in the RG flow as a consequence of the SU(2) spin rotation symmetry \cite{Reuther2010}. 
For the slightly less symmetric Kitaev interactions, only the components $\Gamma^{00}$, $\Gamma^{11}$, $\Gamma^{22}$, and $\Gamma^{33}$ can become non-zero, but unlike in the Heisenberg model, the last three components no longer need to be equal. 

The different parametrizations have an immediate impact on the computational complexity of the problem. 
To leading order, the complexity depends on the number of diagrams that include a summation over the entire lattice (c.f. Fig.~\ref{fig:pffrg:flowequation}). Estimates for the computational complexity for a number of model systems are compared in Table \ref{tab:pffrg:complexity}. 
\begin{table}[t]
	\begin{tabular}{l r}
	\toprule
	Model & rel. complexity \\ 
	\colrule
	Heisenberg interactions & 1 \\ 
	XYZ interactions & 2 \\ 
	Off-diagonal interactions & 32 \\ 
	Time-reversal breaking & 2048 \\
	\botrule
	\end{tabular}
	\caption{{\bf Computational complexity} for different parametrizations of the pf-FRG flow equations. 
	Apart from the parametrization, the complexity also depends on the size of the lattice and the underlying Matsubara frequency mesh (see text for details).
	The parametrization for general off-diagonal interactions proposed in this article is two orders of magnitude simpler than the unconstrained flow equations (that allow breaking of time-reversal symmetry). }
	\label{tab:pffrg:complexity}
\end{table}
As long as time-reversal symmetry remains intact, one may exploit the four symmetries in the frequency dependence listed in Eq. \eqref{eq:pffrg:parametrization4point:1} to cut the computational costs by a factor of 16. 
Additional complexity may arise depending on how many lattice symmetries can be exploited: While Heisenberg interactions usually are not bond dependent, this is a common feature in models with Kitaev-like interactions. DM interactions, in particular, break inversion symmetry on lattice bonds which automatically reduces the lattice's point group. 
In total, the overall computational complexity scales as \[\mathcal{O}(N_L^2 N_\omega^4)\eqs,\] where $N_L$ is the number of symmetry-reduced lattice sites and $N_\omega$ is the number of frequencies that are being used to model the Matsubara frequency dependence. This scaling arises from the necessity to compute $\mathcal{O}(N_L N_\omega^3)$ diagrams, each containing an internal sum over $\mathcal{O}(N_L)$ lattice sites and $\mathcal{O}(N_\omega)$ frequencies.


\section{Symmetry classification}
\label{sec:symmetries} 
We now proceed to a detailed analysis of the (projective) symmetries of the general pseudo-fermion Hamiltonian, which we 
use to derive symmetry constraints on the functional form of single-particle and two-particle correlation functions. These results
in turn allow us to implement a symmetry-constrained parametrization for the effective action \eqref{eq:pffrg:parametrization2point}-\eqref{eq:pffrg:parametrization4point:1} in the pf-FRG scheme in the subsequent sections.
 
To recapitulate, the Hamiltonian for the auxiliary pseudo-fermion degrees of freedom is obtained from the original spin model \eqref{eq:pffrg:hamiltonian} by applying the pseudo-fermion transformation \eqref{eq:pffrg:pseudo-fermions}, and generally reads as
\begin{equation}
\label{eq:symmetries:hamiltonian}
H=\sum\limits_{ij} \frac{J_{ij}^{\mu\nu}}{4} \sigma^\mu_{\alpha\beta} \sigma^\nu_{\gamma\delta} ~f^\dagger_{i\alpha}f^\dagger_{j\gamma}f^\nodagger_{j\delta}f^\nodagger_{i\beta} \eqs.
\end{equation}
This Hamiltonian exhibits two distinct types of symmetries, which should be carefully distinguished:
On the one hand, there are physical symmetries present in the original spin Hamiltonian, most importantly time reversal symmetry which inverts the sign of each spin operators -- but as we are only considering two-spin interactions, these signs cancel. The Hamiltonian is also assumed to be hermitian. Since the spin operators are already hermitian themselves, this limits our analysis to real couplings constants. 

On the other hand, the Hamiltonian has an additional, non-physical symmetry that derives from the fermionization process and is therefore present in any pseudo-fermion Hamiltonian. 
This extra symmetry is a local SU(2) gauge redundancy, an artifact of the parton construction \eqref{eq:pffrg:pseudo-fermions} 
\begin{equation}
\label{eq:symmetries:pseudo-fermions}
S_i^\mu \rightarrow \frac{1}{2}f^\dagger_{i\alpha} \sigma_{\alpha\beta}^\mu f^\nodagger_{i\beta} \eqs. 
\end{equation}
In this notation it is easy to see that there is an inherent U(1) gauge redundancy in the construction which amounts to locally multiplying fermionic operators with an arbitrary phase factor. Since the fermion operators always come in pairs, the phase factor cancels.  
It is less obvious, however, that the real symmetry group is larger. Therefore, let us make the full SU(2) symmetry more explicit by re-writing the above substitution rule. 
Instead of expressing the spin operator in terms of a vector-matrix-vector product, it can also be expressed in terms of a trace over a matrix-matrix-matrix product \cite{Affleck1988} 
\begin{equation}
\label{eq:symmetries:pseudo-fermions:1}
S_i^\mu \rightarrow \frac{1}{4} F_{i,\alpha\beta}^\dagger \sigma^\mu_{\beta\gamma} F_{i,\gamma\alpha}^\nodagger \eqs,
\end{equation}
where the $2\times2$ matrix $F_i$ of pseudo-fermionic operators is defined as
\begin{equation}
F_i = \left( \begin{array}{cc}
f_{i\uparrow}^\nodagger & f_{i\downarrow}^\dagger \\ 
f_{i\downarrow}^\nodagger & -f_{i\uparrow}^\dagger
\end{array} \right) \eqs.
\end{equation}
The local SU(2) gauge redundancy is represented by the space of $2\times2$ matrices $g_\mathrm{local}$ with the defining property $g_\mathrm{local}^\dagger g_\mathrm{local}^\nodagger = 1$, which is the conventional representation spanned by the Pauli matrices. 
The symmetry group action on the pseudo-fermionic operators is given by right-multiplication of $g_\mathrm{local}$ with the operators, 
\begin{equation}
\tilde{F}_i=F_i~g_\mathrm{local} \eqs.
\end{equation}
In this notation, the invariance of the parton construction \eqref{eq:symmetries:pseudo-fermions:1} is a direct consequence of the invariance of the trace under cyclic permutations. 

Moreover, the difference between the artificial SU(2) gauge redundancy and the physical SU(2) spin rotation can be made apparent as well. 
While we have implemented the (local) gauge redundancy as right-multiplication, the (global) physical spin rotation $g_\mathrm{global}$ is implemented as left-multiplication, 
\begin{equation}
\tilde{F}_i=g_\mathrm{global}~F_i \eqs, 
\end{equation}
such that in the spin operator \eqref{eq:symmetries:pseudo-fermions:1} it does not cancel out and instead acts as a rotation on the Pauli matrix spin space, 
\begin{equation}
\tilde{\sigma}^\mu=g_\mathrm{global}^\dagger \sigma^\mu g_\mathrm{global}^\nodagger \eqs,
\end{equation} as expected. 
In the remainder of this section, however, we shall not address the full SU(2) redundancy in one blow, but instead separately consider  the U(1) sub-group and a particle-hole symmetry. 
The reason that we treat the two symmetries separately is that the first one puts a strong constraint on the spatial structure of our parametrization of the vertex functions, while the latter one is used to derive constraints on the frequency structure of the parametrization. 

For each symmetry (regardless of whether it is physical or unphysical), we shall present an implementation of the symmetry group action on the pseudo-fermion space in second quantized language. 
We then derive constraints that the symmetry places on pseudo-fermionic correlation functions. 
Since the structure of the correlation functions is intimately tied to the structure of the single-particle irreducible vertices (in which the pf-FRG scheme is formulated) by construction \cite{Kopietz2010a},
the constraints ultimately carry over to the parametrization of the interaction vertices. 


\subsection{Local U(1) symmetry} 
One of the most important symmetries, that sets pseudo-fermion models apart from conventional fermion systems, is the artificial local U(1) symmetry, a sub-group of the artificial SU(2) gauge redundancy. 
In pseudo-fermion space, we can define the action of local U(1) rotations by a set of angles \{$\varphi_i$\}, where each angle is associated with a lattice site $i$ (not to be confused with the imaginary unit), acting as
\begin{equation}
\label{eq:symmetries:U(1)}
g_{\varphi_i} \left( \begin{array}{c}
f^\dagger_{i\alpha} \\ 
f^\nodagger_{i\alpha}
\end{array} \right) g_{\varphi_i}^{-1} = \left( \begin{array}{c}
e^{i\varphi_{i}} f^\dagger_{i\alpha} \\ 
e^{-i\varphi_i} f^\nodagger_{i\alpha}
\end{array} \right) \eqs. 
\end{equation}
We are now interested in the transformation behavior of single-particle correlation functions
\begin{equation}
G(1' ; 1) = \int \mathrm{d}\tau'\mathrm{d}\tau e^{i\tau'\omega' - i\tau\omega} \left< f^\dagger_{i'\tau'\alpha'} f^\nodagger_{i\tau\alpha} \right> \eqs,
\end{equation}
as well as in the transformation behavior of two-particle correlation functions
\begin{align}
&G(1',2';1,2) \nonumber\\
&\quad= \int \mathrm{d}\tau_{1'}\mathrm{d}\tau_{2'}\mathrm{d}\tau_{1}\mathrm{d}\tau_{2} e^{i(\tau_{1'}\omega_{1'}+\tau_{2'}\omega_{2'} - \tau_{1}\omega_{1} - \tau_{2}\omega_{2})} \nonumber\\
&\quad\quad \times \left< f^\dagger_{i_{1'}\tau_{1'}\alpha_{1'}}f^\dagger_{i_{2'}\tau_{2'}\alpha_{2'}} f^\nodagger_{i_{1}\tau_{1}\alpha_{1}} f^\nodagger_{i_{2}\tau_{2}\alpha_{2}} \right> \eqs,
\end{align}
where we have used the shorthand notation $n={i_n,\omega_n,\alpha_n}$ for composite lattice site, frequency, and spin indices. We suppress the time ordering operator in the correlator, as it becomes trivial once we upgrade pseudo-fermionic operators to Grassmann numbers in the field-theoretical framework that the pf-FRG approach is formulated in. 

In order for the symmetry transformation to leave the correlators invariant, lattice site indices may only appear pairwise in creation and annihilation operators, such that the phase factors vanish. 
For the single-particle correlation function, this poses the constraint 
\begin{equation}
\label{eq:symmetries:U(1):1particle}
G(1' ; 1) = G(1' ; 1)\delta_{i_{1'}i_{1}} \eqs, 
\end{equation}
and for the two-particle correlator we have 
\begin{align}
\label{eq:symmetries:U(1):2particle}
G(1', 2'; 1, 2) &= 
G(1', 2'; 1, 2) \delta_{i_1' i_1}\delta_{i_2' i_2} \nonumber\\
&\quad - G(2', 1'; 1, 2) \delta_{i_2' i_1}\delta_{i_1' i_2} \eqs.
\end{align}
Further details are presented in Appendix \ref{sec:appendix:symmetry}. 

\subsection{Local particle-hole symmetry} 
Next, we consider the artificial local particle-hole symmetry that is a subset of the SU(2) gauge redundancy of the pseudo-fermion Hamiltonian. 
In the pseudo-fermionic space, we define the symmetry operation as 
\begin{equation}
\label{eq:symmetries:PH}
g_i \left( \begin{array}{c}
f^\dagger_{i\alpha} \\ 
f^\nodagger_{i\alpha}
\end{array} \right) g_i^{-1} = \left( \begin{array}{c}
\alpha f^\nodagger_{i\bar{\alpha}} \\ 
\alpha f^\dagger_{i\bar{\alpha}}
\end{array} \right) \eqs,
\end{equation}
where the spin index $\alpha$ takes values $+1$ or $-1$ (representing spin-up and spin-down, respectively). The notation $\bar{\alpha}$ indicates that the spin has been reversed, $\bar{\alpha}=-\alpha$. 
This transformation leaves the pseudo-fermion Hamiltonian \eqref{eq:symmetries:hamiltonian} invariant and, requiring that the single-particle correlation functions also remains invariant, yields the relation
\begin{equation}
\label{eq:symmetries:PH:1particle}
G(1';1) = -\alpha'\alpha G(i -\omega \bar{\alpha} ; i' -\omega' \bar{\alpha}')  \eqs.
\end{equation}
Note that for conciseness we are omitting commas in between triples of lattice site index, Matsubara frequency, and spin index; The expression $G(i -\omega \bar{\alpha} ; i' -\omega' \bar{\alpha}')$ should therefore be read as $G(i, -\omega, \bar{\alpha} ; i', -\omega', \bar{\alpha}')$. 
On the level of bi-local two-particle correlators (we have learned from the local U(1) symmetry that we only need to consider bi-local correlation functions), we obtain two different symmetry relations since we can independently apply the particle-hole transformation on the two lattice sites: 
\begin{align}
\label{eq:symmetries:PH:2particle}
&G(1',2';1,2)\delta_{i_{1'}i_{1}}\delta_{i_{2'}i_{2}} \nonumber\\
&\quad = -\alpha_{1'}\alpha_{1} G(i_{1}-\omega_{1}\bar{\alpha}_{1},i_{2}\omega_{2'}\alpha_{2'};i_{1}-\omega_{1'}\bar{\alpha}_{1'},i_{2}\omega_{2}\alpha_{2}) \nonumber\\
&\quad = -\alpha_{2'}\alpha_{2} G(i_{1}\omega_{1'}\alpha_{1'},i_{2}-\omega_{2}\bar{\alpha}_{2};i_{1}\omega_{1}\alpha_{1},i_{2}-\omega_{2'}\bar{\alpha}_{2'}) ~.
\end{align}
In fact, the symmetries also hold independently for the purely local vertex $i_1=i_2$, since the particle-hole transformation is inherently tied to the pseudo-fermion construction.
More precisely, it acts on pairs of fermions that originate from the same spin operator in the fermionization process \eqref{eq:symmetries:pseudo-fermions}. 
Since these pairs of fermions necessarily live on the same lattice site, it is often simpler to think of the symmetry as a local transformation. 
For a more rigorous treatment one would need to introduce additional indices that carry the information of which spin operator each  individual fermion belongs to, and define the particle-hole symmetry to act locally in this extra index space. 
As such a treatment does not generate new insight, we shall refrain from writing it down explicitly and simply impose that the symmetry relations \eqref{eq:symmetries:PH:2particle} also hold for purely local correlators.


\subsection{Lattice symmetries} 
\label{sec:symmetries:lattice}
We now focus on a second set of symmetries that affects the structure of lattice site indices -- genuine lattice symmetries.
They are necessarily present in the microscopic definition of the spin models that we consider. 
Depending on the specifics of the microscopic model and the type of lattice that it is defined on, the group of lattice symmetries varies in its number of symmetry elements. 
Any lattice can be defined via an underlying periodic Bravais lattice, which is decorated with a single- or multi-atomic unit cell. 
In lattice calculations, it is often convenient to group lattice sites by their relative position in the unit cell, such that every sublattice individually preserves the translation symmetry of the Bravais lattice. 
In pf-FRG calculations, we typically do not discriminate between different sublattices, but assume all lattice sites to be identical, i.e. we assume that it is possible to map any lattice site to any other site via a lattice symmetry. 
Such transformations exist also for non-Bravais lattices, but they may require more complex transformations that go beyond plain translations. 

The straight-forward definition of a lattice transformation only acts on the lattice site index,  
\begin{equation}
\label{eq:symmetries:Lattice}
g_{T} \left( \begin{array}{c}
f^\dagger_{i\alpha} \\ 
f^\nodagger_{i\alpha}
\end{array} \right)g_{T}^{-1} = \left( \begin{array}{c}
f^\dagger_{T(i)\alpha} \\ 
f^\nodagger_{T(i)\alpha}
\end{array} \right) \eqs, 
\end{equation}
where $T$ is a lattice automorphism, which maps the lattice onto itself. 
This definition is sufficient for most Heisenberg-like spin models, but it could also be upgraded to a combined symmetry in lattice space and spin space, which can be particularly useful in the presence of bond-directional interactions, as defined, for instance, in the Kitaev honeycomb model. 
For the single-particle correlation function, this implies 
\begin{equation}
\label{eq:symmetries:Lattice:1particle}
G(1' ; 1) = G(T(i') \omega'\alpha' ; T(i) \omega\alpha) \eqs.
\end{equation}
For the two-particle correlator, it implies
\begin{align}
\label{eq:symmetries:Lattice:2particle}
G(1',2';1,2) =G\big( &T(i_{1'})\omega_{1'}\alpha_{1'},T(i_{2'}) \omega_{2'}\alpha_{2'} ; \nonumber\\
&T(i_{1}) \omega_{1}\alpha_{1}, T(i_{2}) \omega_{2}\alpha_{2} \big)  \eqs.
\end{align}
Using the locality constraint for single-particle correlator \eqref{eq:symmetries:U(1):1particle}, its site dependence can be reduced from two lattice sites to a single lattice site. In combination with lattice symmetries, the single site dependence can be further mapped to an arbitrary (fixed) reference site, and in our notation we may suppress the site dependence altogether. 
The lattice site structure of the two-particle correlator, which is a function of two lattice sites as a consequence of the bi-locality constraint \eqref{eq:symmetries:U(1):2particle}, can be further reduced to depend only on a single site (in addition to a fixed reference site).


\subsection{Time-reversal symmetry}
We now proceed to the physical symmetries of the Hamiltonian and first examine time-reversal symmetry. Time-reversal is an anti-unitary symmetry which effectively inverts the sign of every spin operator. Consequently, all two-spin interactions with real coupling constants, which are captured by the general Hamiltonian \eqref{eq:pffrg:hamiltonian}, preserve time-reversal symmetry. 
On the Hilbert space of pseudo-fermions, the symmetry operation can be implemented as 
\begin{equation}
\label{eq:symmetries:TR}
g \left( \begin{array}{c}
f^\dagger_{i\alpha} \\ 
f^\nodagger_{i\alpha}
\end{array} \right) g^{-1} = \left( \begin{array}{c}
e^{i\pi \alpha/2} f^\dagger_{i\bar{\alpha}} \\ 
e^{-i\pi \alpha/2} f^\nodagger_{i\bar{\alpha}}
\end{array} \right) \eqs,
\end{equation}
where $g$ is anti-unitary. 
Note that this definition is not unique, since it can always be composed with arbitrary transformations from the SU(2) gauge redundancy. 
Analyzing its effects on single-particle correlators and bi-local two-particle correlators, we obtain the symmetry relations 
\begin{equation}
\label{eq:symmetries:TR:1particle}
G(1';1) = \alpha'\alpha G(i' -\omega' \bar{\alpha}' ; i -\omega \bar{\alpha})^* 
\end{equation}
and
\begin{align}
\label{eq:symmetries:TR:2particle}
&G(1',2';1,2)\delta_{i_{1'}i_{1}}\delta_{i_{2'}i_{2}} \nonumber\\
&\quad = \alpha_{1'}\alpha_{2'}\alpha_{1}\alpha_{2} G(i_{1}-\omega_{1'}\bar{\alpha}_{1'},i_{2}-\omega_{2'}\bar{\alpha}_{2'}; \nonumber\\
&\quad\quad\quad i_{1}-\omega_{1}\bar{\alpha}_{1},i_{2}-\omega_{2}\bar{\alpha}_{2})^* \eqs,
\end{align}
respectively. 
The anti-unitary property of the transformation introduces a complex conjugation, which we can exploit to make a connection between the real and imaginary parts of the correlators. 


\subsection{Hermitian symmetry}
The relations that we have derived in the presence of time-reversal symmetry become a lot more powerful in combination with a hermitian symmetry, i.e. assuming that the Hamiltonian is self-adjoint. 
The two-spin Hamiltonian \eqref{eq:pffrg:hamiltonian} automatically fulfills this condition, since the individual spin operators are already self-adjoint and we assume all prefactors to be real. 
Therefore, we may upgrade the complex conjugation in relations \eqref{eq:symmetries:TR:1particle} and \eqref{eq:symmetries:TR:2particle} to a conjugate transpose (since they are plain numbers, transposition acts trivially) and evaluate the expressions explicitly. 
Leaving the Hamiltonian -- and hence the Boltzmann factors -- in the thermal expectation value invariant, the constraints on the correlation functions can be evaluated to 
\begin{equation}
\label{eq:symmetries:H:1particle}
G(1';1) = \alpha'\alpha G(i \omega \bar{\alpha} ; i' \omega' \bar{\alpha}')
\end{equation}
for the single-particle correlator and 
\begin{align}
\label{eq:symmetries:H:2particle}
&G(1',2';1,2)\delta_{i_{1'}i_{1}}\delta_{i_{2'}i_{2}}  = \alpha_{1'}\alpha_{2'}\alpha_{1}\alpha_{2} \nonumber\\
&\quad\times G(i_{1}\omega_{1}\bar{\alpha}_{1},i_{2}\omega_{2}\bar{\alpha}_{2};i_{1}\omega_{1'}\bar{\alpha}_{1'},i_{2}\omega_{2'}\bar{\alpha}_{2'})
\end{align}
for the bi-local two-particle correlator. 


\section{Symmetry-constrained vertex parametrization}
\label{sec:VertexParametrization}

The key result of the detailed analysis of the individual symmetries of the general bilinear spin interactions of form \eqref{eq:pffrg:hamiltonian} in the previous Section has been to derive constraints on the functional form of single-particle and two-particle correlation functions, summarized with regard to the individual symmetries in the final equations of each of its subsections.
With these symmetry considerations in place, we now proceed to combine these individual symmetries to find a convenient parametrization for the correlation functions that ultimately leads us to an efficient, symmetry-constrained parametrization of the effective action \eqref{eq:pffrg:parametrization2point}-\eqref{eq:pffrg:parametrization4point:1} in the pf-FRG scheme. 

\begin{table*}[t]
\begin{align}
\tag{H $\circ$ TR}
G^\mu(\omega) &= \xi(\mu)G^\mu(\omega) \\
\tag{PH}
G^\mu(\omega) &= -\xi(\mu)G^\mu(-\omega) \\
\tag{TR $\circ$ PH}
G^\mu(\omega) &= -G^\mu(\omega)^* \nonumber\\
\tag{X $\circ$ H $\circ$ TR $\circ$ PH1 $\circ$ PH2}
G^{\mu\nu}_{i_1i_2}(s,t,u)&= G^{\nu\mu}_{i_2i_1}(-s,t,u) \\
\tag{H $\circ$ TR}
G^{\mu\nu}_{i_1i_2}(s,t,u)&= \xi(\mu)\xi(\nu)G^{\mu\nu}_{i_1i_2}(s,-t,u) \\
\tag{X $\circ$ H $\circ$ TR}
G^{\mu\nu}_{i_1i_2}(s,t,u)&= \xi(\mu)\xi(\nu) G^{\nu\mu}_{i_2i_1}(s,t,-u) \\
\tag{PH2}
G^{\mu\nu}_{i_1i_2}(s,t,u)&= -\xi(\nu) G^{\mu\nu}_{i_1i_2}(u,t,s) \\
\tag{TR $\circ$ H $\circ$ TR $\circ$ PH1 $\circ$ PH2}
G^{\mu\nu}_{i_1i_2}(s,t,u)&= \xi(\mu)\xi(\nu) G^{\mu\nu}_{i_1i_2}(s,t,u)^* 
\end{align}
\caption{{\bf Symmetry constraints on the basis functions} for the parametrization of one-particle and two-particle vertices in the pf-FRG scheme. The equations are labeled by the symmetries that have been used to derive them -- `H' is shorthand notation for the hermitian symmetry, `TR' is time reversal, `X' denotes the simultaneous exchange of the two in-going and the two out-going fermion operators, and `PH1' and `PH2' are the two particle-hole symmetries acting on the first pair of lattice sites and the second pair, respectively (if applied to a single-particle vertex, this distinction is not sensible).}
\label{tab:parametrization:basisfunctions}
\end{table*}

Let us begin with the parametrization of the single-particle correlation function. 
Starting from the general expression for the single-particle correlator $G(1';1)$, we use the local U(1) symmetry to guarantee locality in real space. In combination with lattice symmetries, we can always map the single lattice site dependence to a fixed reference site. The correlation function thereby becomes independent of the lattice site and we suppress the site index in our notation. 
Furthermore, we shall make use of Matsubara frequency conservation (as a consequence of translation symmetry in imaginary time, not shown explicitly) to see that the correlation function must be diagonal in the frequency index. 
The remaining dependency on spin indices is captured by an expansion in the basis of Pauli matrices, such that the correlator can generally be written as 
\begin{equation}
G(1';1) = \left( G^\mu(\omega) \sigma^\mu_{\alpha'\alpha} \right)\delta_{i'i}\delta_{\omega'\omega} \eqs,
\end{equation}
where we are implicitly summing over the repeated index $\mu=0,\dots,3$. The $2\times2$ matrix $\sigma^0$ denotes the identity matrix and $\sigma^1$, $\sigma^2$, and $\sigma^3$ are the Pauli matrices. 

For the parametrization of the two-particle vertex, we proceed analogously. Utilizing the local U(1) symmetry in combination with lattice symmetries and Matsubara frequency conservation, we can conveniently write the correlator as
\begin{align}
\label{eq:parametrization:parametrization4point}
&G(1', 2'; 1, 2) = \nonumber\\
&\quad\left[ \left( G^{\mu\nu}_{i_1i_2}(s,t,u) \sigma^\mu_{\alpha_{1'}\alpha_{1}} \sigma^\nu_{\alpha_{2'}\alpha_{2}} \right) \delta_{i_{1'} i_1}\delta_{i_{2'} i_2} - (1' \leftrightarrow 2' ) \right] \nonumber\\
&\quad\times\delta_{\omega_{1'}+\omega_{2'}-\omega_{1}-\omega_{2}} \eqs, 
\end{align}
where we introduced the transfer frequencies
\begin{align}
s &= \omega_{1'}+\omega_{2'} \nonumber\\
t &= \omega_{1'}-\omega_{1} \nonumber\\
u &= \omega_{1'}-\omega_{2} \eqs. 
\end{align}

Using the set of symmetry relations, which we have explicitly derived in the subsections of the previous Section, we find that the basis functions for the single-particle correlator, $G^\mu(\omega)$, and for the two-particle correlator, $G^{\mu\nu}_{i_1i_2}(s,t,u)$, are constrained by the relations given in Table \ref{tab:parametrization:basisfunctions}. 
In these relations, we have introduced the sign function 
\begin{equation}
\xi(\mu)= \left\{ \begin{array}{rl}
+1 &\mbox{if $\mu=0$} \\
-1 &\mbox{otherwise} \end{array} \right.
\end{equation}
that results from symmetry manipulations of spin indices after using the identities
\begin{equation}
\alpha'\alpha \sigma^\mu_{\bar{\alpha}\bar{\alpha}'} = \alpha'\alpha \left( \sigma^{\mu *} \right)_{\bar{\alpha}'\bar{\alpha}} = \xi(\mu) \sigma^\mu_{\alpha'\alpha} \eqs. 
\end{equation}

With this, we have almost reached the goal to find an efficient, symmetry-constrained parametrization for the effective action in the pf-FRG scheme. 
The one step left is to see that the symmetries from the (disconnected) correlation functions carry over to the effective action -- that is, to the one-line irreducible correlation functions $\Sigma(1';1)$ and $\Gamma(1',2';1,2)$. 
For the single-particle correlation function this is easy to see, since per definition the relation 
\begin{equation}
G(1';1)=\frac{1}{i\omega-\Sigma(1';1)} 
\end{equation}
must hold. If $G$ is diagonal in all arguments and anti-symmetric the frequency dependence, this must also be true for $\Sigma$. 
For the two-particle correlation function, it is not as easy to see that the symmetries carry over. 
But we can see the inheritance of symmetries from the so-called tree-expansion \cite{Kopietz2010a}, which relates the one-line irreducible two-particle correlation function $\Gamma(1',2';1,2)$ to the connected correlation function $G_c(1',2';1,2)$ according to 
\begin{align}
G_c(1',2';1,2) &= -\sum\limits_{3456} \Gamma(3,4;5,6) \nonumber\\
&\quad \times G(1';3) G(2';4) G(5;1) G(6;2) \eqs,
\end{align}
which has a structure simple enough for the symmetries to directly carry over (knowing that $G$ is diagonal in all its arguments). 
Between the connected correlation function $G_c$ and disconnected correlation function $G$, it can then be seen on the level of their generating functionals that they have the same symmetries \cite{Kopietz2010a}. 

We can thus conclude that the parametrization \eqref{eq:pffrg:parametrization2point}-\eqref{eq:pffrg:parametrization4point:1} of the effective action is valid for any time-reversal symmetric, hermitian pseudo-fermion Hamiltonian. 
In particular, it is valid for any two-spin interaction with real coupling constants.


\section{Symmetries of the  flow equations }
\label{sec:symmetriesFlowEquation}
 
By considering symmetries of the Hamiltonian, we have demonstrated, in the previous Section, that the effective action can be efficiently parametrized by a set of purely real functions, c.f. Eqs. \eqref{eq:pffrg:parametrization2point} -- \eqref{eq:pffrg:parametrization4point:1}. 
We now complement these findings by arguing that, on the level of the pf-FRG flow equations, the parametrization and the symmetries of its basis functions are indeed preserved throughout the renormalization group flow -- assuming that they exist in the initial conditions. 
Starting from the familiar parametrization of the self-energy, 
\begin{equation}
\Sigma(1';1) = \Sigma(\omega_{1}) \delta_{\alpha_{1'}\alpha_{1}}\delta_{i_{1'}i_{1}}\delta_{\omega_{1'}\omega_{1}} \eqs, 
\end{equation}
with
\begin{equation}
\Sigma(\omega_{1})\in i\mathbb{R} \eqs,
\end{equation}
and the parametrization of the two-particle vertex, 
\begin{align}
&\Gamma(1', 2'; 1, 2) = \nonumber\\
&\quad\left[ \left( \Gamma^{\mu\nu}_{i_1i_2}(s,t,u) \sigma^\mu_{\alpha_{1'}\alpha_{1}} \sigma^\nu_{\alpha_{2'}\alpha_{2}} \right) \delta_{i_{1'} i_1}\delta_{i_{2'} i_2} - (1' \leftrightarrow 2' ) \right] \nonumber\\
&\quad\times\delta_{\omega_{1'}+\omega_{2'}-\omega_{1}-\omega_{2}} \eqs,
\end{align}
with 
\begin{equation}
\Gamma^{\mu\nu}_{i_1i_2}(s,t,u) \in 
\left\{ \begin{array}{l}
\phantom i\mathbb{R} \quad \mbox{if $\xi(\mu)\xi(\nu)=1$} \\
i\mathbb{R} \quad \mbox{if $\xi(\mu)\xi(\nu)=-1$} 
\end{array} \right. \eqs,
\end{equation}
it can readily be seen, by inserting the expressions into the FRG flow equations \eqref{eq:pffrg:flow:1particle} and \eqref{eq:pffrg:flow:2particle}, that the parametrization is complete and no additional terms are generated throughout the RG flow. 

On top of the parametrization, we postulate the following symmetry relations
\begin{align}
\label{eq:symmetriesFlowEquation}
&\Sigma(\omega_{1})=-\Sigma(-\omega_{1}) \nonumber\\
&\Gamma^\Lambda(1',2';1,2) = \Gamma^\Lambda(2',1';2,1) \nonumber\\
&\Gamma^\Lambda(1',2';1,2) = \Gamma^\Lambda(1,2;1',2')^* \nonumber\\
&\Gamma^\Lambda(1',2';1,2) = \nonumber\\
&\quad\Gamma^\Lambda(i_{1'}-\omega_{1'}\alpha_{1'},i_{2'}-\omega_{2'}\alpha_{2'};i_{1}-\omega_{1}\alpha_{1},i_{2}-\omega_{2}\alpha_{2}) \nonumber\\
&\Gamma_{i_1i_2}^\Lambda(1',2';1,2) =\nonumber\\
&\quad -\alpha_{2'}\alpha_2 \Gamma_{i_1i_2}^\Lambda(\omega_{1'}\alpha_{1'},-\omega_{2}\bar{\alpha}_{2};\omega_{1}\alpha_{1},-\omega_{2'}\bar{\alpha}_{2'}) \eqs.
\end{align}
The first two symmetry relations have already been discussed in the previous Section. 
The third relation is a combination of Eqs. \eqref{eq:symmetries:PH:2particle} and \eqref{eq:symmetries:TR:2particle}. 
The fourth relation is a combination of Eqs. \eqref{eq:symmetries:PH:2particle} and \eqref{eq:symmetries:H:2particle}. 
The last relation is equivalent to Eq. \eqref{eq:symmetries:PH:2particle}. 
By combining these symmetries with the vertex parametrizations, one recovers the symmetry constraints on the basis functions as listed in Eq. \eqref{eq:pffrg:parametrization4point:1}. 

In this form, however, the symmetries are more convenient to verify on the level of the flow equations. 
This is done by inserting the symmetry relations \eqref{eq:symmetriesFlowEquation} into the FRG flow equations, given in Eqs. \eqref{eq:pffrg:flow:1particle} and \eqref{eq:pffrg:flow:2particle}, and confirming that the derivative of the vertices has the same symmetry as the vertices themselves -- the explicit calculation is provided in Appendix \ref{sec:appendix:induction}. 
This insight proves, by induction, that the symmetries are preserved throughout the \emph{entire} RG flow.


\section{Application to kagome magnets}
\label{sec:application}
To illustrate the numerical efficiency of the symmetry-constrained pf-FRG scheme introduced in the discussion of the previous Sections, we apply it to the spin-1/2 Heisenberg antiferromagnet on the kagome lattice augmented by a general (in-plane and out-of-plance) Dzyaloshinskii-Moriya (DM) exchange. This general form of an off-diagonal DM interaction requires to make full use of the symmetry constraints introduced above and was previously beyond the numerical scope of the pf-FRG approach.

\begin{figure}[t]
\includegraphics[width=0.7\linewidth]{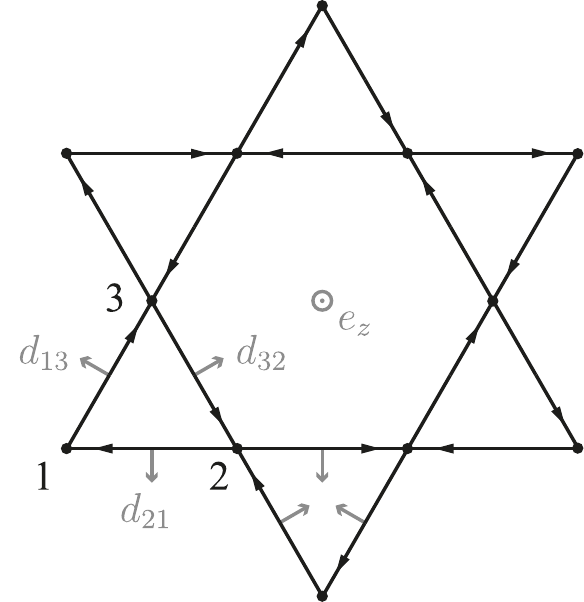}
\caption{{\bf DM interactions on the kagome lattice.} On each lattice bond, the orientation of the DM coupling $(D \mathbf{e}_z + D' \mathbf{d}_{ij}) \cdot \left( \mathbf{S}_i \times \mathbf{S}_j \right)$ is defined by the black arrows pointing from site $i$ to $j$. The orientation of the in-plane component of the DM vectors is different for up-pointing and down-pointing triangles, as indicated by the gray arrows. The vectors $\mathbf{e}_z$ and $\mathbf{d}_{ij}$ have unit length. }
\label{fig:application:model}
\end{figure}

Explicitly, the microscopical model of interest is captured by the Hamiltonian 
\begin{equation}
\label{eq:application:model}
\sum\limits_{\langle i,j\rangle} J~\mathbf{S}_i \cdot \mathbf{S}_j + (D \mathbf{e}_z + D' \mathbf{d}_{ij}) \cdot \left( \mathbf{S}_i \times \mathbf{S}_j \right) \eqs, 
\end{equation}
where the DM vectors $D \mathbf{e}_z + D' \mathbf{d}_{ij}$ have an out-of-plane component $D$ and an in-plane component $D'$ whose orientation  \footnote{{For our pf-FRG calculations, it is convenient to rotate the lattice such that it lies in the (111) plane. Correspondingly, when we refer to  $\langle S^z S^z \rangle$ correlations, we compute them in (111) direction.}} is defined according to Fig.~\ref{fig:application:model}. 
The strength of the antiferromagnetic Heisenberg interaction in given by a positive $J>0$.
The physical motivation to study Hamiltonian \eqref{eq:application:model} originates, for instance, from the microscopics of the 
spin liquid candidate material \emph{herbertsmithite} \cite{Norman2016}. For this material, it has been   argued that the dominant antiferromagnetic nearest-neighbor interaction of Heisenberg type is accompanied by a sub-dominant DM interaction, which indeed exhibits in-plane and out-of-place components \cite{Zorko2008,Rigol2007}, as schematically illustrated in Fig.~\ref{fig:application:model}.

In our numerical pf-FRG calculations for this model system, we typically consider a finite lattice geometry that extends seven bond lengths in every direction, and correlations are truncated beyond this range. 
Note that such a truncation scheme does not introduce an artificial finite-size boundary to the lattice, but it can rather be understood similar to a series expansion that eventually converges, upon increasing the cutoff range, to the thermodynamic value  \cite{Buessen2018}. 
We make use of the full set of lattice symmetries, including non-trivial symmetries that allow us to map any lattice site onto an arbitrary fixed reference site as detailed in Sec.~\ref{sec:symmetries:lattice}; Note that, despite the model explicitly breaking inversion symmetry of the lattice, mappings between different basis sits on the lattice are still possible by augmenting lattice transformations with rotations in spin space. 
We model the frequency dependence by a set of discrete frequencies arranged symmetrically around zero and interpolate linearly in between the mesh points. In our calculations, we use between $N_\omega=66$ and $N_\omega=144$ frequency points on a logarithmic scale with positive values lying in the range between $\omega_\mathrm{min}=0.001$ and $\omega_\mathrm{max}=250$. 
Employing the symmetry relations listed in Eq.~\ref{eq:pffrg:parametrization4point:1}, it is sufficient to consider only positive frequencies, resulting in a total number of up to $3.7\cdot 10^8$ coupled differential equations that need to be solved per set of coupling constants. 
The flow equations are reasonable well-behaved, such that they can be solved by means of the Euler scheme. We approximate the cutoff dependence by an exponential mesh with a step size $\Lambda_{n+1}=0.98 \Lambda_n$ in the range of $\Lambda_\mathrm{min}=0.01$ and $\Lambda_\mathrm{max}=500$. 

\begin{figure}
\includegraphics[width=0.95\linewidth]{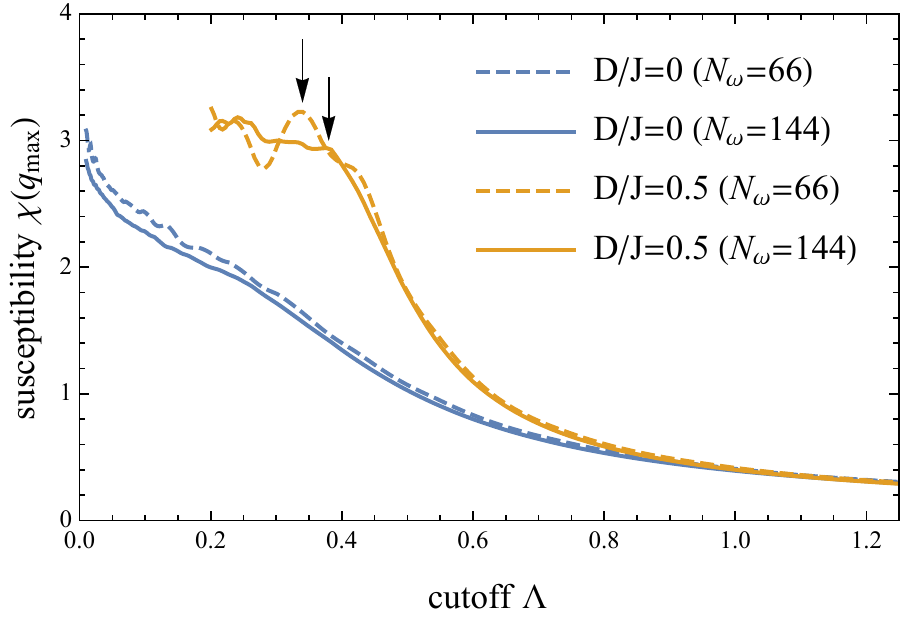}
\caption{{\bf Breakdown scale} of the smooth renormalization group flow. The susceptibility is plotted at the point $q_\mathrm{max}$ in momentum space, where it is largest. A phase transition into a magnetically ordered state is indicated by a breakdown of the smooth flow (black arrows). The breakdown scale is resolved sufficiently well at $N_\omega=144$. Lower resolution of the frequency mesh, $N_\omega=66$, introduces additional numerical uncertainty, which manifests in the form of oscillations on top of the susceptibility flow, making the precise determination of the breakdown scale more difficult. }
\label{fig:application:flowComparison}
\end{figure}
\begin{figure}
\includegraphics[width=0.95\linewidth]{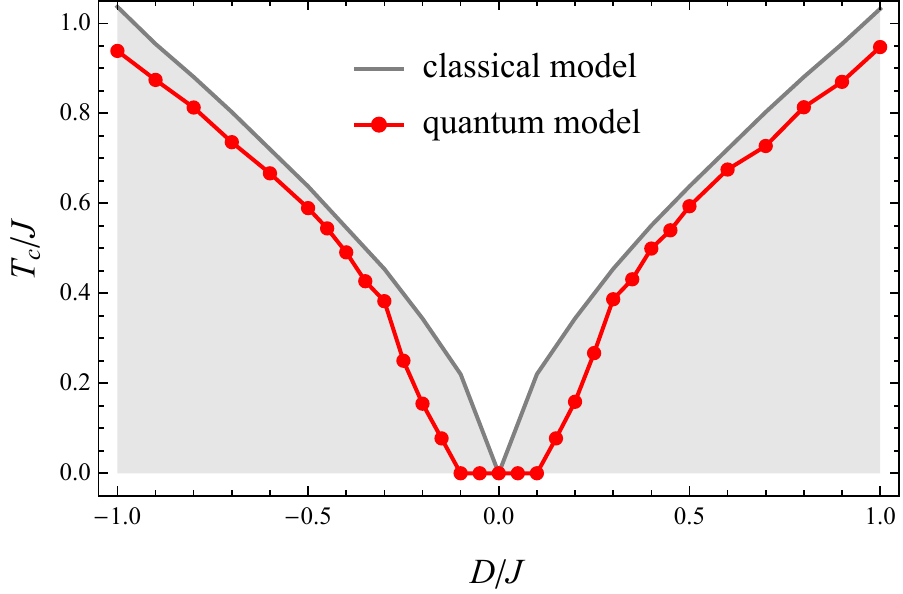}
\caption{{\bf Transition temperature} into the ordered state as a function of the out-of-plane DM coupling $D$ (in-plane coupling $D'$ is set to zero). In the quantum model, a stable non-magnetic phase exists (the kagome antiferromagnet), that is not present in the classical model. The classical results are taken from Ref. \cite{Elhajal2002}.}
\label{fig:application:transitionTemperature}
\end{figure}
\begin{figure}
\includegraphics[width=\linewidth]{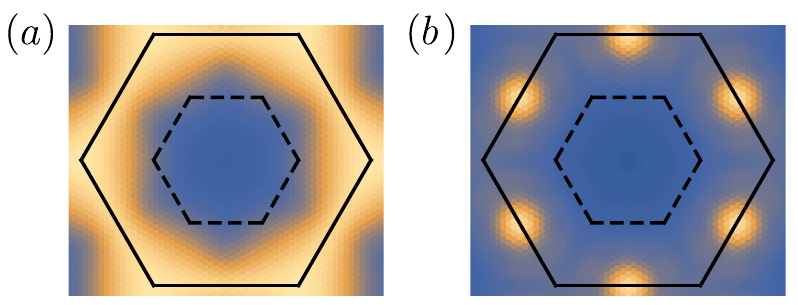}
\caption{{\bf Structure factors} (a) in the spin liquid phase at $D/J=0$ and (b) in the magnetically ordered phase at $D/J=1$ (in-plane coupling $D'$ is set to zero in both figures). The solid black line denotes the extended Brillouin zone, the dashed line indicates the first Brillouin zone.}
\label{fig:application:structurefactor}
\end{figure}

We first consider the case where the in-plane component of the DM interaction vanishes, i.e.~$D'=0$.
It has been established in previous studies \cite{Hering2017} that the spin liquid state, which nucleates around the unperturbed kagome Heisenberg antiferromagnet (KHAFM), remains stable under weak out-of-plane DM interactions up to approximately $D/J\approx 0.1$. 
Our calculations for this scenario with purely out-of-plane DM interactions, summarized in Figs.~\ref{fig:application:flowComparison}, \ref{fig:application:transitionTemperature}, and \ref{fig:application:structurefactor}, confirm these findings. 
As shown in Fig.~\ref{fig:application:flowComparison}, we find that below the critical coupling a smooth evolution of the susceptibility flow is found down to zero cutoff, indicating the existence of a low-temperature spin liquid regime. 
If the DM coupling exceeds the critical value, a breakdown of the smooth flow indicates the onset of spontaneous symmetry breaking, and the system undergoes a magnetic ordering transition.
From the breakdown scale, we can extract an estimate for the transition temperature $T_c=\frac{\pi}{2}\Lambda_c$ \cite{Iqbal2016} and determine the finite-temperature phase diagram of the model, see Fig.~\ref{fig:application:transitionTemperature}.
While the static structure factor  
\begin{equation}
\chi(\mathbf{q}) = \frac{1}{N^2} \sum\limits_{i,j} \mathrm{e}^{i \mathbf{q} (\mathrm{r}_i - \mathrm{r}_j)} \langle S_i^z S_j^z \rangle \eqs,
\end{equation}
is featureless for the KHAFM (i.e.~for $D=0$) and shows no signs of magnetic order, the structure factor for finite out-of-plane DM interactions shows clear maxima at positions that are associated with $q=0$ order, see the two panels of Fig.~\ref{fig:application:structurefactor}. 
Away from the spin liquid regime, we therefore find the same type of magnetic order that is known to proliferate in the classical model \cite{Elhajal2002}. 
The transition temperature, as compared to the classical model, is slightly lowered in the presence of quantum fluctuations,
as documented by the direct comparison in Fig.~\ref{fig:application:transitionTemperature}. 

\begin{figure}[t]
\includegraphics[width=0.95\linewidth]{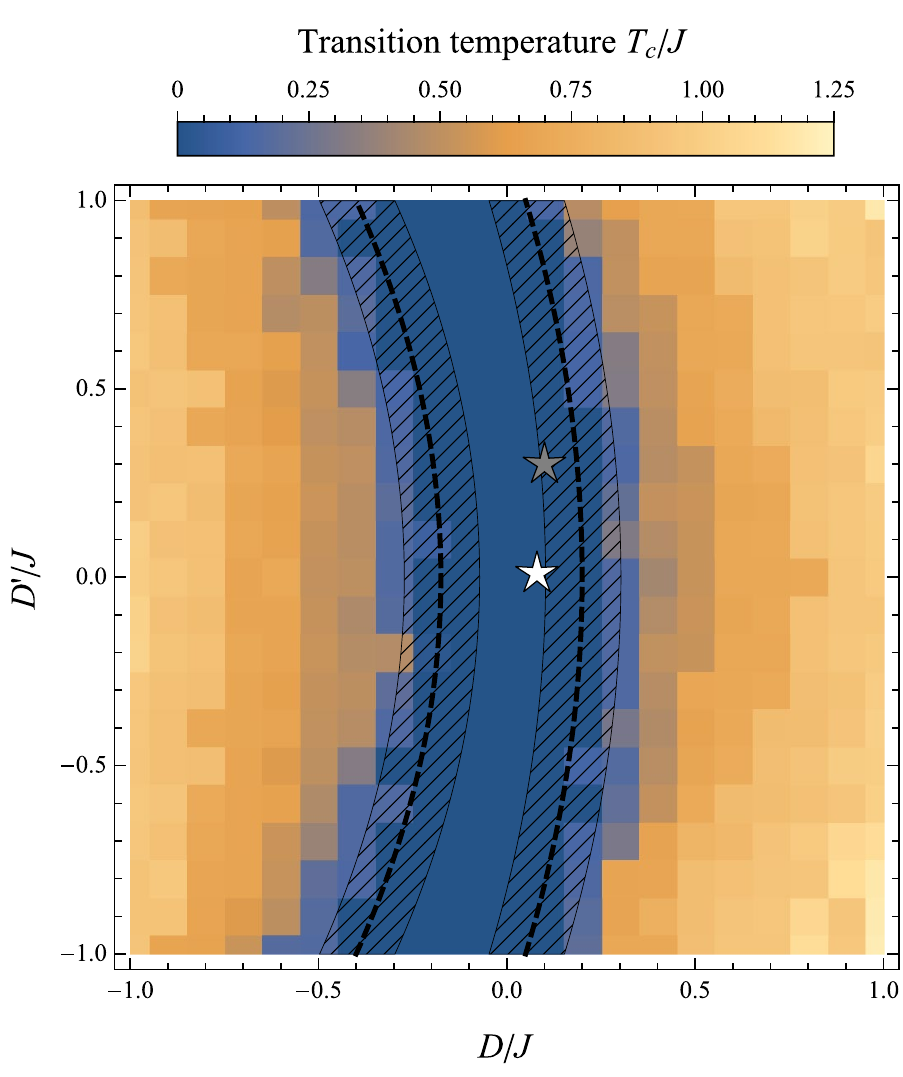}
\caption{{\bf Phase diagram} of the full Hamiltonian with out-of-plane DM interaction $D$ and in-plane DM interaction $D'$. Blue color indicates the region where no magnetic ordering transition is observed down to zero temperature. The approximate phase boundaries are indicated by the dashed lines. The frequency resolution is $N_\omega=66$, which leads to an error bar of the phase boundary of approximately $D/J \pm 0.1$, as indicated by the shaded region (typically, lower frequency resolution in pf-FRG tends to overestimate paramagnetic regions, c.f. Fig.~\ref{fig:application:flowComparison}). The white and gray stars indicate estimates of the coupling constants in herbertsmithite as determined in Refs. \cite{Zorko2008} and \cite{Rigol2007}, respectively. } 
\label{fig:application:phaseDiagram}
\end{figure}

As we tune the in-plane DM interaction to finite values $D'/J>0$, we can again ask about the stability of the spin liquid phase. We find that the spin liquid indeed persists even when $D'$ is of similar strength as the Heisenberg coupling. 
However, once we include also out-of-plane DM interactions, the precise location of the phase boundary between the spin liquid phase and the magnetically ordered phase is shifted depending on the value of $D'$, see Fig.~\ref{fig:application:phaseDiagram} for a complete ground state phase diagram as a function of the in-plane and out-of-plane DM coupling strengths $D'$ and $D$, respectively. 
Initially, the phase boundary is symmetric around $D=0$, but at finite in-plane components $|D'|>0$ the transition points are shifted towards smaller $D$. 
Such a bending of the phase boundaries has already been seen in the classical model, where the spin liquid phase is not present. Instead, there is a direct transition between two magnetically ordered phases of $q=0$ type that differ in their chirality \cite{Elhajal2002}. 
Unfortunately, the direct measurement of the spin chirality  $\chi_{ijk}=\mathbf{S}_i \cdot ( \mathbf{S}_j \times \mathbf{S}_k ) $, where $i$, $j$, and $k$ are sites on an elementary triangle, is not possible in pf-FRG calculations, as it breaks time reversal symmetry
and involves three spin operators. 

For herbertsmithite, the strength of the DM interactions has been estimated in electron spin resonance measurements \cite{Zorko2008} to be $D/J \approx 0.08$ and $D'/J \approx 0.01$. Other model calculations that focus on reproducing the thermodynamic properties of the material report similar out-of-plane DM interactions, but more sizable in-plane interactions of up to $D'/J \approx 0.3$ \cite{Rigol2007}.
Regardless of the actual size of the in-plane couplings, our calculations imply that any finite in-plane DM interaction pushes the system closer towards the ordered $q=0$ state and hence is compatible with the weak maxima that have been measured in  inelastic neutron scattering measurements \cite{Han2016} of the structure factor at points that are associated with $q=0$ order. 


\section{Conclusions}
\label{sec:conclusions}
To summarize, we have generalized the pf-FRG method such that arbitrary spin-anisotropic models with two-body exchange couplings may be efficiently handled. Particularly, our approach allows to numerically treat off-diagonal $\Gamma$ interactions that have been discussed in the context of various Kitaev materials and general Dzyaloshinskii-Moriya exchanges even in the case where no continuous spin-rotation symmetries are present. The main difficultly of this generalization concerns the appearance of new spin components of the fermionic two-particle vertex and, as a consequence, an enormous growth of the complexity of the RG flow equations. We have demonstrated, based on a detailed symmetry analysis and an efficient parametrization of the vertex functions, that the complexity can be limited to a degree that only leads to moderate increase of the computational costs as compared to spin isotropic systems. Key simplifications of the RG equations are achieved by exploiting combinations of time reversal symmetry and an SU(2) gauge redundancy that is intimately connected to the pseudo-fermionic representation of the original spin operators. Due to a subtle interplay of these properties, the fermionic self-energy assumes a simple diagonal form in its spin variables and the two-particle vertex satisfies various symmetries in its frequency arguments resulting in an overall drastic reduction of numerical complexity.

As a first demonstration of its capabilities we have applied our generalized pf-FRG approach to an anisotropic spin system on the kagome lattice with nearest neighbor Heisenberg exchange $J$ as well as in-plane and out-of-plane Dzyaloshinskii-Moriya interactions $D'$ and $D$, respectively. Our main finding is that the well-known non-magnetic phase of the $J$-only model is readily destabilized into $q=0$ magnetic order by rather small out-of-plane DM interactions $D/J\approx0.1$, while the inclusion of in-plane DM components leaves this phase largely intact.

It is worth highlighting again that our pf-FRG algorithm is applicable to general anisotropic two-body spin interactions and, hence, provides a flexible methodological framework for the investigation of an abundance of spin systems. For example, the currently investigated Kitaev candidate materials have been proposed to harbor the full range of symmetry-allowed exchange couplings including Kitaev, $\Gamma$, and Dzyaloshinskii-Moriya interactions which now become amenable to a numerical pf-FRG analysis. Another possible future research direction are anisotropic spin interactions for pyrochlore quantum magnets, which exhibit a rich phenomenology such as the emergent electrodynamics in quantum spin ice systems. Ultimately, even though computationally costly, it will also be worth further expanding the scope of the pf-FRG by including finite magnetic fields which, e.g., will allow one to investigate field induced quantum spin liquids.


\begin{acknowledgments}
V.N. and J.R. thank M. Hering for discussions and ongoing collaborations on related projects.
We acknowledge partial support from  the Deutsche Forschungsgemeinschaft (DFG, German Research Foundation), Projektnummer 277146847 -- SFB 1238 (project C03) and Projektnummer 277101999 -- TRR 183 (project A02).
The numerical simulations were performed on the CHEOPS cluster at RRZK Cologne and the JURECA Booster at the Forschungszentrum Juelich. 
\end{acknowledgments}


\bibliography{triFRG}

\begin{thebibliography}{45}%
\makeatletter
\providecommand \@ifxundefined [1]{%
 \@ifx{#1\undefined}
}%
\providecommand \@ifnum [1]{%
 \ifnum #1\expandafter \@firstoftwo
 \else \expandafter \@secondoftwo
 \fi
}%
\providecommand \@ifx [1]{%
 \ifx #1\expandafter \@firstoftwo
 \else \expandafter \@secondoftwo
 \fi
}%
\providecommand \natexlab [1]{#1}%
\providecommand \enquote  [1]{``#1''}%
\providecommand \bibnamefont  [1]{#1}%
\providecommand \bibfnamefont [1]{#1}%
\providecommand \citenamefont [1]{#1}%
\providecommand \href@noop [0]{\@secondoftwo}%
\providecommand \href [0]{\begingroup \@sanitize@url \@href}%
\providecommand \@href[1]{\@@startlink{#1}\@@href}%
\providecommand \@@href[1]{\endgroup#1\@@endlink}%
\providecommand \@sanitize@url [0]{\catcode `\\12\catcode `\$12\catcode
  `\&12\catcode `\#12\catcode `\^12\catcode `\_12\catcode `\%12\relax}%
\providecommand \@@startlink[1]{}%
\providecommand \@@endlink[0]{}%
\providecommand \url  [0]{\begingroup\@sanitize@url \@url }%
\providecommand \@url [1]{\endgroup\@href {#1}{\urlprefix }}%
\providecommand \urlprefix  [0]{URL }%
\providecommand \Eprint [0]{\href }%
\providecommand \doibase [0]{http://dx.doi.org/}%
\providecommand \selectlanguage [0]{\@gobble}%
\providecommand \bibinfo  [0]{\@secondoftwo}%
\providecommand \bibfield  [0]{\@secondoftwo}%
\providecommand \translation [1]{[#1]}%
\providecommand \BibitemOpen [0]{}%
\providecommand \bibitemStop [0]{}%
\providecommand \bibitemNoStop [0]{.\EOS\space}%
\providecommand \EOS [0]{\spacefactor3000\relax}%
\providecommand \BibitemShut  [1]{\csname bibitem#1\endcsname}%
\let\auto@bib@innerbib\@empty
\bibitem [{\citenamefont {{F.~D.~M.
  Haldane}}(1983{\natexlab{a}})}]{Haldane1983a}%
  \BibitemOpen
  \bibfield  {author} {\bibinfo {author} {\bibnamefont {{F.~D.~M. Haldane}}},\
  }\bibfield  {title} {{\color{Gray}\small \bibinfo {title} {{Continuum
  dynamics of the 1-D Heisenberg antiferromagnet: Identification with the O(3)
  nonlinear sigma model}},\ }}\href {\doibase
  https://doi.org/10.1016/0375-9601(83)90631-X} {\bibfield  {journal} {\bibinfo
   {journal} {Physics Letters A}\ }\textbf {\bibinfo {volume} {93}},\ \bibinfo
  {pages} {464 } (\bibinfo {year} {1983}{\natexlab{a}})}\BibitemShut {NoStop}%
\bibitem [{\citenamefont {{F.~D.~M.
  Haldane}}(1983{\natexlab{b}})}]{Haldane1983b}%
  \BibitemOpen
  \bibfield  {author} {\bibinfo {author} {\bibnamefont {{F.~D.~M. Haldane}}},\
  }\bibfield  {title} {{\color{Gray}\small \bibinfo {title} {{Nonlinear Field
  Theory of Large-Spin Heisenberg Antiferromagnets: Semiclassically Quantized
  Solitons of the One-Dimensional Easy-Axis N\'eel State}},\ }}\href {\doibase
  10.1103/PhysRevLett.50.1153} {\bibfield  {journal} {\bibinfo  {journal}
  {Phys. Rev. Lett.}\ }\textbf {\bibinfo {volume} {50}},\ \bibinfo {pages}
  {1153} (\bibinfo {year} {1983}{\natexlab{b}})}\BibitemShut {NoStop}%
\bibitem [{\citenamefont {Gu}\ and\ \citenamefont {Wen}(2009)}]{SPT}%
  \BibitemOpen
  \bibfield  {author} {\bibinfo {author} {\bibfnamefont {Z.-C.}\ \bibnamefont
  {Gu}}\ and\ \bibinfo {author} {\bibfnamefont {X.-G.}\ \bibnamefont {Wen}},\
  }\bibfield  {title} {{\color{Gray}\small \bibinfo {title}
  {{Tensor-entanglement-filtering renormalization approach and
  symmetry-protected topological order}},\ }}\href {\doibase
  10.1103/PhysRevB.80.155131} {\bibfield  {journal} {\bibinfo  {journal} {Phys.
  Rev. B}\ }\textbf {\bibinfo {volume} {80}},\ \bibinfo {pages} {155131}
  (\bibinfo {year} {2009})}\BibitemShut {NoStop}%
\bibitem [{\citenamefont {Lieb}\ \emph {et~al.}(1961)\citenamefont {Lieb},
  \citenamefont {Schultz},\ and\ \citenamefont {Mattis}}]{LSM}%
  \BibitemOpen
  \bibfield  {author} {\bibinfo {author} {\bibfnamefont {E.}~\bibnamefont
  {Lieb}}, \bibinfo {author} {\bibfnamefont {T.}~\bibnamefont {Schultz}}, \
  and\ \bibinfo {author} {\bibfnamefont {D.}~\bibnamefont {Mattis}},\
  }\bibfield  {title} {{\color{Gray}\small \bibinfo {title} {Two soluble models
  of an antiferromagnetic chain},\ }}\href {\doibase
  https://doi.org/10.1016/0003-4916(61)90115-4} {\bibfield  {journal} {\bibinfo
   {journal} {Annals of Physics}\ }\textbf {\bibinfo {volume} {16}},\ \bibinfo
  {pages} {407} (\bibinfo {year} {1961})}\BibitemShut {NoStop}%
\bibitem [{\citenamefont {Oshikawa}(2000)}]{Oshikawa2000}%
  \BibitemOpen
  \bibfield  {author} {\bibinfo {author} {\bibfnamefont {M.}~\bibnamefont
  {Oshikawa}},\ }\bibfield  {title} {{\color{Gray}\small \bibinfo {title}
  {{Commensurability, Excitation Gap, and Topology in Quantum Many-Particle
  Systems on a Periodic Lattice}},\ }}\href {\doibase
  10.1103/PhysRevLett.84.1535} {\bibfield  {journal} {\bibinfo  {journal}
  {Phys. Rev. Lett.}\ }\textbf {\bibinfo {volume} {84}},\ \bibinfo {pages}
  {1535} (\bibinfo {year} {2000})}\BibitemShut {NoStop}%
\bibitem [{\citenamefont {Hastings}(2004)}]{Hastings2004}%
  \BibitemOpen
  \bibfield  {author} {\bibinfo {author} {\bibfnamefont {M.~B.}\ \bibnamefont
  {Hastings}},\ }\bibfield  {title} {{\color{Gray}\small \bibinfo {title}
  {{Lieb-Schultz-Mattis in higher dimensions}},\ }}\href {\doibase
  10.1103/PhysRevB.69.104431} {\bibfield  {journal} {\bibinfo  {journal} {Phys.
  Rev. B}\ }\textbf {\bibinfo {volume} {69}},\ \bibinfo {pages} {104431}
  (\bibinfo {year} {2004})}\BibitemShut {NoStop}%
\bibitem [{\citenamefont {Chen}\ \emph {et~al.}(2010)\citenamefont {Chen},
  \citenamefont {Gu},\ and\ \citenamefont {Wen}}]{Chen2010}%
  \BibitemOpen
  \bibfield  {author} {\bibinfo {author} {\bibfnamefont {X.}~\bibnamefont
  {Chen}}, \bibinfo {author} {\bibfnamefont {Z.-C.}\ \bibnamefont {Gu}}, \ and\
  \bibinfo {author} {\bibfnamefont {X.-G.}\ \bibnamefont {Wen}},\ }\bibfield
  {title} {{\color{Gray}\small \bibinfo {title} {Local unitary transformation,
  long-range quantum entanglement, wave function renormalization, and
  topological order},\ }}\href {\doibase 10.1103/PhysRevB.82.155138} {\bibfield
   {journal} {\bibinfo  {journal} {Phys. Rev. B}\ }\textbf {\bibinfo {volume}
  {82}},\ \bibinfo {pages} {155138} (\bibinfo {year} {2010})}\BibitemShut
  {NoStop}%
\bibitem [{\citenamefont {Levin}\ and\ \citenamefont
  {Wen}(2006)}]{LevinWen2006}%
  \BibitemOpen
  \bibfield  {author} {\bibinfo {author} {\bibfnamefont {M.}~\bibnamefont
  {Levin}}\ and\ \bibinfo {author} {\bibfnamefont {X.-G.}\ \bibnamefont
  {Wen}},\ }\bibfield  {title} {{\color{Gray}\small \bibinfo {title}
  {{Detecting Topological Order in a Ground State Wave Function}},\ }}\href
  {\doibase 10.1103/PhysRevLett.96.110405} {\bibfield  {journal} {\bibinfo
  {journal} {Phys. Rev. Lett.}\ }\textbf {\bibinfo {volume} {96}},\ \bibinfo
  {pages} {110405} (\bibinfo {year} {2006})}\BibitemShut {NoStop}%
\bibitem [{\citenamefont {Kitaev}\ and\ \citenamefont
  {Preskill}(2006)}]{KitaevPreskill2006}%
  \BibitemOpen
  \bibfield  {author} {\bibinfo {author} {\bibfnamefont {A.}~\bibnamefont
  {Kitaev}}\ and\ \bibinfo {author} {\bibfnamefont {J.}~\bibnamefont
  {Preskill}},\ }\bibfield  {title} {{\color{Gray}\small \bibinfo {title}
  {{Topological Entanglement Entropy}},\ }}\href {\doibase
  10.1103/PhysRevLett.96.110404} {\bibfield  {journal} {\bibinfo  {journal}
  {Phys. Rev. Lett.}\ }\textbf {\bibinfo {volume} {96}},\ \bibinfo {pages}
  {110404} (\bibinfo {year} {2006})}\BibitemShut {NoStop}%
\bibitem [{\citenamefont {Savary}\ and\ \citenamefont
  {Balents}(2017)}]{Savary2016}%
  \BibitemOpen
  \bibfield  {author} {\bibinfo {author} {\bibfnamefont {L.}~\bibnamefont
  {Savary}}\ and\ \bibinfo {author} {\bibfnamefont {L.}~\bibnamefont
  {Balents}},\ }\bibfield  {title} {{\color{Gray}\small \bibinfo {title}
  {Quantum spin liquids: a review},\ }}\href
  {http://stacks.iop.org/0034-4885/80/i=1/a=016502} {\bibfield  {journal}
  {\bibinfo  {journal} {Reports on Progress in Physics}\ }\textbf {\bibinfo
  {volume} {80}},\ \bibinfo {pages} {016502} (\bibinfo {year}
  {2017})}\BibitemShut {NoStop}%
\bibitem [{\citenamefont {Wen}(2002)}]{Wen2002}%
  \BibitemOpen
  \bibfield  {author} {\bibinfo {author} {\bibfnamefont {X.-G.}\ \bibnamefont
  {Wen}},\ }\bibfield  {title} {{\color{Gray}\small \bibinfo {title} {{Quantum
  orders and symmetric spin liquids}},\ }}\href {\doibase
  10.1103/PhysRevB.65.165113} {\bibfield  {journal} {\bibinfo  {journal} {Phys.
  Rev. B}\ }\textbf {\bibinfo {volume} {65}},\ \bibinfo {pages} {165113}
  (\bibinfo {year} {2002})}\BibitemShut {NoStop}%
\bibitem [{\citenamefont {Bednorz}\ and\ \citenamefont
  {M{\"u}ller}(1986)}]{Bednorz1986}%
  \BibitemOpen
  \bibfield  {author} {\bibinfo {author} {\bibfnamefont {J.~G.}\ \bibnamefont
  {Bednorz}}\ and\ \bibinfo {author} {\bibfnamefont {K.~A.}\ \bibnamefont
  {M{\"u}ller}},\ }\bibfield  {title} {{\color{Gray}\small \bibinfo {title}
  {{Possible high$T_c$ superconductivity in the Ba-La-Cu-O system}},\ }}\href
  {\doibase 10.1007/BF01303701} {\bibfield  {journal} {\bibinfo  {journal}
  {Zeitschrift f{\"u}r Physik B Condensed Matter}\ }\textbf {\bibinfo {volume}
  {64}},\ \bibinfo {pages} {189} (\bibinfo {year} {1986})}\BibitemShut
  {NoStop}%
\bibitem [{\citenamefont {Auerbach}(1998)}]{Auerbach1998}%
  \BibitemOpen
  \bibfield  {author} {\bibinfo {author} {\bibfnamefont {A.}~\bibnamefont
  {Auerbach}},\ }\href@noop {} {\emph {\bibinfo {title} {{Interacting Electrons
  and Quantum Magnetism}}}},\ Graduate Texts in Contemporary Physics\ (\bibinfo
   {publisher} {Springer New York},\ \bibinfo {year} {1998})\BibitemShut
  {NoStop}%
\bibitem [{\citenamefont {Kitaev}(2006)}]{Kitaev2006}%
  \BibitemOpen
  \bibfield  {author} {\bibinfo {author} {\bibfnamefont {A.}~\bibnamefont
  {Kitaev}},\ }\bibfield  {title} {{\color{Gray}\small \bibinfo {title} {Anyons
  in an exactly solved model and beyond},\ }}\href {\doibase
  http://dx.doi.org/10.1016/j.aop.2005.10.005} {\bibfield  {journal} {\bibinfo
  {journal} {Annals of Physics}\ }\textbf {\bibinfo {volume} {321}},\ \bibinfo
  {pages} {2 } (\bibinfo {year} {2006})}\BibitemShut {NoStop}%
\bibitem [{\citenamefont {Trebst}()}]{Trebst2017}%
  \BibitemOpen
  \bibfield  {author} {\bibinfo {author} {\bibfnamefont {S.}~\bibnamefont
  {Trebst}},\ }\bibfield  {title} {{\color{Gray}\small \bibinfo {title}
  {{Kitaev Materials}},\ }}\href@noop {} {\ }\Eprint
  {http://arxiv.org/abs/arXiv:1701.07056} {arXiv:1701.07056} \BibitemShut
  {NoStop}%
\bibitem [{\citenamefont {Jackeli}\ and\ \citenamefont
  {Khaliullin}(2009)}]{Jackeli2009}%
  \BibitemOpen
  \bibfield  {author} {\bibinfo {author} {\bibfnamefont {G.}~\bibnamefont
  {Jackeli}}\ and\ \bibinfo {author} {\bibfnamefont {G.}~\bibnamefont
  {Khaliullin}},\ }\bibfield  {title} {{\color{Gray}\small \bibinfo {title}
  {{Mott Insulators in the Strong Spin-Orbit Coupling Limit: From Heisenberg to
  a Quantum Compass and Kitaev Models}},\ }}\href {\doibase
  10.1103/PhysRevLett.102.017205} {\bibfield  {journal} {\bibinfo  {journal}
  {Phys. Rev. Lett.}\ }\textbf {\bibinfo {volume} {102}},\ \bibinfo {pages}
  {017205} (\bibinfo {year} {2009})}\BibitemShut {NoStop}%
\bibitem [{\citenamefont {Witczak-Krempa}\ \emph {et~al.}(2014)\citenamefont
  {Witczak-Krempa}, \citenamefont {Chen}, \citenamefont {Kim},\ and\
  \citenamefont {Balents}}]{WitczakKrempa2014}%
  \BibitemOpen
  \bibfield  {author} {\bibinfo {author} {\bibfnamefont {W.}~\bibnamefont
  {Witczak-Krempa}}, \bibinfo {author} {\bibfnamefont {G.}~\bibnamefont
  {Chen}}, \bibinfo {author} {\bibfnamefont {Y.~B.}\ \bibnamefont {Kim}}, \
  and\ \bibinfo {author} {\bibfnamefont {L.}~\bibnamefont {Balents}},\
  }\bibfield  {title} {{\color{Gray}\small \bibinfo {title} {{Correlated
  Quantum Phenomena in the Strong Spin-Orbit Regime}},\ }}\href {\doibase
  10.1146/annurev-conmatphys-020911-125138} {\bibfield  {journal} {\bibinfo
  {journal} {Annual Review of Condensed Matter Physics}\ }\textbf {\bibinfo
  {volume} {5}},\ \bibinfo {pages} {57} (\bibinfo {year} {2014})}\BibitemShut
  {NoStop}%
\bibitem [{\citenamefont {Rau}\ \emph {et~al.}(2016)\citenamefont {Rau},
  \citenamefont {Lee},\ and\ \citenamefont {Kee}}]{Rau2016}%
  \BibitemOpen
  \bibfield  {author} {\bibinfo {author} {\bibfnamefont {J.~G.}\ \bibnamefont
  {Rau}}, \bibinfo {author} {\bibfnamefont {E.~K.-H.}\ \bibnamefont {Lee}}, \
  and\ \bibinfo {author} {\bibfnamefont {H.-Y.}\ \bibnamefont {Kee}},\
  }\bibfield  {title} {{\color{Gray}\small \bibinfo {title} {{Spin-Orbit
  Physics Giving Rise to Novel Phases in Correlated Systems: Iridates and
  Related Materials}},\ }}\href {\doibase
  10.1146/annurev-conmatphys-031115-011319} {\bibfield  {journal} {\bibinfo
  {journal} {Annual Review of Condensed Matter Physics}\ }\textbf {\bibinfo
  {volume} {7}},\ \bibinfo {pages} {195} (\bibinfo {year} {2016})}\BibitemShut
  {NoStop}%
\bibitem [{\citenamefont {Winter}\ \emph {et~al.}(2016)\citenamefont {Winter},
  \citenamefont {Li}, \citenamefont {Jeschke},\ and\ \citenamefont
  {Valent\'{\i}}}]{Winter2016}%
  \BibitemOpen
  \bibfield  {author} {\bibinfo {author} {\bibfnamefont {S.~M.}\ \bibnamefont
  {Winter}}, \bibinfo {author} {\bibfnamefont {Y.}~\bibnamefont {Li}}, \bibinfo
  {author} {\bibfnamefont {H.~O.}\ \bibnamefont {Jeschke}}, \ and\ \bibinfo
  {author} {\bibfnamefont {R.}~\bibnamefont {Valent\'{\i}}},\ }\bibfield
  {title} {{\color{Gray}\small \bibinfo {title} {{Challenges in design of
  Kitaev materials: Magnetic interactions from competing energy scales}},\
  }}\href {\doibase 10.1103/PhysRevB.93.214431} {\bibfield  {journal} {\bibinfo
   {journal} {Phys. Rev. B}\ }\textbf {\bibinfo {volume} {93}},\ \bibinfo
  {pages} {214431} (\bibinfo {year} {2016})}\BibitemShut {NoStop}%
\bibitem [{\citenamefont {Reuther}\ and\ \citenamefont
  {W{\"{o}}lfle}(2010)}]{Reuther2010}%
  \BibitemOpen
  \bibfield  {author} {\bibinfo {author} {\bibfnamefont {J.}~\bibnamefont
  {Reuther}}\ and\ \bibinfo {author} {\bibfnamefont {P.}~\bibnamefont
  {W{\"{o}}lfle}},\ }\bibfield  {title} {{\color{Gray}\small \bibinfo {title}
  {{$J_1$-$J_2$ frustrated two-dimensional Heisenberg model: Random phase
  approximation and functional renormalization group}},\ }}\href {\doibase
  10.1103/PhysRevB.81.144410} {\bibfield  {journal} {\bibinfo  {journal} {Phys.
  Rev. B}\ }\textbf {\bibinfo {volume} {81}},\ \bibinfo {pages} {144410}
  (\bibinfo {year} {2010})}\BibitemShut {NoStop}%
\bibitem [{\citenamefont {Wetterich}(1993)}]{Wetterich1993}%
  \BibitemOpen
  \bibfield  {author} {\bibinfo {author} {\bibfnamefont {C.}~\bibnamefont
  {Wetterich}},\ }\bibfield  {title} {{\color{Gray}\small \bibinfo {title}
  {{Exact evolution equation for the effective potential}},\ }}\href {\doibase
  10.1016/0370-2693(93)90726-X} {\bibfield  {journal} {\bibinfo  {journal}
  {Physics Letters B}\ }\textbf {\bibinfo {volume} {301}},\ \bibinfo {pages}
  {90} (\bibinfo {year} {1993})}\BibitemShut {NoStop}%
\bibitem [{\citenamefont {Reuther}\ \emph {et~al.}(2011)\citenamefont
  {Reuther}, \citenamefont {Thomale},\ and\ \citenamefont
  {Trebst}}]{Reuther2011c}%
  \BibitemOpen
  \bibfield  {author} {\bibinfo {author} {\bibfnamefont {J.}~\bibnamefont
  {Reuther}}, \bibinfo {author} {\bibfnamefont {R.}~\bibnamefont {Thomale}}, \
  and\ \bibinfo {author} {\bibfnamefont {S.}~\bibnamefont {Trebst}},\
  }\bibfield  {title} {{\color{Gray}\small \bibinfo {title}
  {{Finite-temperature phase diagram of the Heisenberg-Kitaev model}},\ }}\href
  {\doibase 10.1103/PhysRevB.84.100406} {\bibfield  {journal} {\bibinfo
  {journal} {Phys. Rev. B}\ }\textbf {\bibinfo {volume} {84}},\ \bibinfo
  {pages} {100406(R)} (\bibinfo {year} {2011})}\BibitemShut {NoStop}%
\bibitem [{\citenamefont {Hering}\ and\ \citenamefont
  {Reuther}(2017)}]{Hering2017}%
  \BibitemOpen
  \bibfield  {author} {\bibinfo {author} {\bibfnamefont {M.}~\bibnamefont
  {Hering}}\ and\ \bibinfo {author} {\bibfnamefont {J.}~\bibnamefont
  {Reuther}},\ }\bibfield  {title} {{\color{Gray}\small \bibinfo {title}
  {{Functional renormalization group analysis of Dzyaloshinsky-Moriya and
  Heisenberg spin interactions on the kagome lattice}},\ }}\href {\doibase
  10.1103/PhysRevB.95.054418} {\bibfield  {journal} {\bibinfo  {journal} {Phys.
  Rev. B}\ }\textbf {\bibinfo {volume} {95}},\ \bibinfo {pages} {054418}
  (\bibinfo {year} {2017})}\BibitemShut {NoStop}%
\bibitem [{\citenamefont {Zorko}\ \emph {et~al.}(2008)\citenamefont {Zorko},
  \citenamefont {Nellutla}, \citenamefont {van Tol}, \citenamefont {Brunel},
  \citenamefont {Bert}, \citenamefont {Duc}, \citenamefont {Trombe},
  \citenamefont {de~Vries}, \citenamefont {Harrison},\ and\ \citenamefont
  {Mendels}}]{Zorko2008}%
  \BibitemOpen
  \bibfield  {author} {\bibinfo {author} {\bibfnamefont {A.}~\bibnamefont
  {Zorko}}, \bibinfo {author} {\bibfnamefont {S.}~\bibnamefont {Nellutla}},
  \bibinfo {author} {\bibfnamefont {J.}~\bibnamefont {van Tol}}, \bibinfo
  {author} {\bibfnamefont {L.~C.}\ \bibnamefont {Brunel}}, \bibinfo {author}
  {\bibfnamefont {F.}~\bibnamefont {Bert}}, \bibinfo {author} {\bibfnamefont
  {F.}~\bibnamefont {Duc}}, \bibinfo {author} {\bibfnamefont {J.-C.}\
  \bibnamefont {Trombe}}, \bibinfo {author} {\bibfnamefont {M.~A.}\
  \bibnamefont {de~Vries}}, \bibinfo {author} {\bibfnamefont {A.}~\bibnamefont
  {Harrison}}, \ and\ \bibinfo {author} {\bibfnamefont {P.}~\bibnamefont
  {Mendels}},\ }\bibfield  {title} {{\color{Gray}\small \bibinfo {title}
  {{Dzyaloshinsky-Moriya Anisotropy in the Spin-1/2 Kagome Compound
  ZnCu3(OH)6Cl2}},\ }}\href {\doibase 10.1103/PhysRevLett.101.026405}
  {\bibfield  {journal} {\bibinfo  {journal} {Phys. Rev. Lett.}\ }\textbf
  {\bibinfo {volume} {101}},\ \bibinfo {pages} {026405} (\bibinfo {year}
  {2008})}\BibitemShut {NoStop}%
\bibitem [{\citenamefont {Rigol}\ and\ \citenamefont
  {Singh}(2007)}]{Rigol2007}%
  \BibitemOpen
  \bibfield  {author} {\bibinfo {author} {\bibfnamefont {M.}~\bibnamefont
  {Rigol}}\ and\ \bibinfo {author} {\bibfnamefont {R.~R.~P.}\ \bibnamefont
  {Singh}},\ }\bibfield  {title} {{\color{Gray}\small \bibinfo {title} {{Kagome
  lattice antiferromagnets and Dzyaloshinsky-Moriya interactions}},\ }}\href
  {\doibase 10.1103/PhysRevB.76.184403} {\bibfield  {journal} {\bibinfo
  {journal} {Phys. Rev. B}\ }\textbf {\bibinfo {volume} {76}},\ \bibinfo
  {pages} {184403} (\bibinfo {year} {2007})}\BibitemShut {NoStop}%
\bibitem [{\citenamefont {Norman}(2016)}]{Norman2016}%
  \BibitemOpen
  \bibfield  {author} {\bibinfo {author} {\bibfnamefont {M.~R.}\ \bibnamefont
  {Norman}},\ }\bibfield  {title} {{\color{Gray}\small \bibinfo {title}
  {{Colloquium: Herbertsmithite and the search for the quantum spin liquid}},\
  }}\href {\doibase 10.1103/RevModPhys.88.041002} {\bibfield  {journal}
  {\bibinfo  {journal} {Rev. Mod. Phys.}\ }\textbf {\bibinfo {volume} {88}},\
  \bibinfo {pages} {041002} (\bibinfo {year} {2016})}\BibitemShut {NoStop}%
\bibitem [{\citenamefont {Suttner}\ \emph {et~al.}(2014)\citenamefont
  {Suttner}, \citenamefont {Platt}, \citenamefont {Reuther},\ and\
  \citenamefont {Thomale}}]{Suttner2014}%
  \BibitemOpen
  \bibfield  {author} {\bibinfo {author} {\bibfnamefont {R.}~\bibnamefont
  {Suttner}}, \bibinfo {author} {\bibfnamefont {C.}~\bibnamefont {Platt}},
  \bibinfo {author} {\bibfnamefont {J.}~\bibnamefont {Reuther}}, \ and\
  \bibinfo {author} {\bibfnamefont {R.}~\bibnamefont {Thomale}},\ }\bibfield
  {title} {{\color{Gray}\small \bibinfo {title} {{Renormalization group
  analysis of competing quantum phases in the ${J}_{1}$-${J}_{2}$ Heisenberg
  model on the kagome lattice}},\ }}\href {\doibase 10.1103/PhysRevB.89.020408}
  {\bibfield  {journal} {\bibinfo  {journal} {Phys. Rev. B}\ }\textbf {\bibinfo
  {volume} {89}},\ \bibinfo {pages} {020408(R)} (\bibinfo {year}
  {2014})}\BibitemShut {NoStop}%
\bibitem [{\citenamefont {Buessen}\ and\ \citenamefont
  {Trebst}(2016)}]{Buessen2016}%
  \BibitemOpen
  \bibfield  {author} {\bibinfo {author} {\bibfnamefont {F.~L.}\ \bibnamefont
  {Buessen}}\ and\ \bibinfo {author} {\bibfnamefont {S.}~\bibnamefont
  {Trebst}},\ }\bibfield  {title} {{\color{Gray}\small \bibinfo {title}
  {{Competing magnetic orders and spin liquids in two- and three-dimensional
  kagome systems: Pseudofermion functional renormalization group
  perspective}},\ }}\href {\doibase 10.1103/PhysRevB.94.235138} {\bibfield
  {journal} {\bibinfo  {journal} {Phys. Rev. B}\ }\textbf {\bibinfo {volume}
  {94}},\ \bibinfo {pages} {235138} (\bibinfo {year} {2016})}\BibitemShut
  {NoStop}%
\bibitem [{\citenamefont {Buessen}()}]{Buessen2019a}%
  \BibitemOpen
  \bibfield  {author} {\bibinfo {author} {\bibfnamefont {F.~L.}\ \bibnamefont
  {Buessen}},\ }\emph {\bibinfo {title} {{A Functional Renormalization Group
  Perspective on Quantum Spin Liquids in Three-Dimensional Frustrated
  Magnets}}},\ \href {https://kups.ub.uni-koeln.de/9986} {Ph.D. thesis},\
  \bibinfo  {school} {University of Cologne}\BibitemShut {NoStop}%
\bibitem [{\citenamefont {Baez}\ and\ \citenamefont
  {Reuther}(2017)}]{Baez2017}%
  \BibitemOpen
  \bibfield  {author} {\bibinfo {author} {\bibfnamefont {M.~L.}\ \bibnamefont
  {Baez}}\ and\ \bibinfo {author} {\bibfnamefont {J.}~\bibnamefont {Reuther}},\
  }\bibfield  {title} {{\color{Gray}\small \bibinfo {title} {{Numerical
  treatment of spin systems with unrestricted spin length S: A functional
  renormalization group study}},\ }}\href {\doibase 10.1103/PhysRevB.96.045144}
  {\bibfield  {journal} {\bibinfo  {journal} {Phys. Rev. B}\ }\textbf {\bibinfo
  {volume} {96}},\ \bibinfo {pages} {045144} (\bibinfo {year}
  {2017})}\BibitemShut {NoStop}%
\bibitem [{\citenamefont {Buessen}\ \emph
  {et~al.}(2018{\natexlab{a}})\citenamefont {Buessen}, \citenamefont {Roscher},
  \citenamefont {Diehl},\ and\ \citenamefont {Trebst}}]{Buessen2018a}%
  \BibitemOpen
  \bibfield  {author} {\bibinfo {author} {\bibfnamefont {F.~L.}\ \bibnamefont
  {Buessen}}, \bibinfo {author} {\bibfnamefont {D.}~\bibnamefont {Roscher}},
  \bibinfo {author} {\bibfnamefont {S.}~\bibnamefont {Diehl}}, \ and\ \bibinfo
  {author} {\bibfnamefont {S.}~\bibnamefont {Trebst}},\ }\bibfield  {title}
  {{\color{Gray}\small \bibinfo {title} {{Functional renormalization group
  approach to SU(N) Heisenberg models: Real-space RG at arbitrary N}},\ }}\href
  {\doibase 10.1103/PhysRevB.97.064415} {\bibfield  {journal} {\bibinfo
  {journal} {Phys. Rev. B}\ }\textbf {\bibinfo {volume} {97}},\ \bibinfo
  {pages} {064415} (\bibinfo {year} {2018}{\natexlab{a}})}\BibitemShut
  {NoStop}%
\bibitem [{\citenamefont {Roscher}\ \emph {et~al.}(2018)\citenamefont
  {Roscher}, \citenamefont {Buessen}, \citenamefont {Scherer}, \citenamefont
  {Trebst},\ and\ \citenamefont {Diehl}}]{Roscher2018}%
  \BibitemOpen
  \bibfield  {author} {\bibinfo {author} {\bibfnamefont {D.}~\bibnamefont
  {Roscher}}, \bibinfo {author} {\bibfnamefont {F.~L.}\ \bibnamefont
  {Buessen}}, \bibinfo {author} {\bibfnamefont {M.~M.}\ \bibnamefont
  {Scherer}}, \bibinfo {author} {\bibfnamefont {S.}~\bibnamefont {Trebst}}, \
  and\ \bibinfo {author} {\bibfnamefont {S.}~\bibnamefont {Diehl}},\ }\bibfield
   {title} {{\color{Gray}\small \bibinfo {title} {{Functional renormalization
  group approach to SU(N) Heisenberg models: Momentum-space RG for the large-N
  limit}},\ }}\href {\doibase 10.1103/PhysRevB.97.064416} {\bibfield  {journal}
  {\bibinfo  {journal} {Phys. Rev. B}\ }\textbf {\bibinfo {volume} {97}},\
  \bibinfo {pages} {064416} (\bibinfo {year} {2018})}\BibitemShut {NoStop}%
\bibitem [{\citenamefont {Iqbal}\ \emph {et~al.}(2016)\citenamefont {Iqbal},
  \citenamefont {Thomale}, \citenamefont {{Parisen Toldin}}, \citenamefont
  {Rachel},\ and\ \citenamefont {Reuther}}]{Iqbal2016}%
  \BibitemOpen
  \bibfield  {author} {\bibinfo {author} {\bibfnamefont {Y.}~\bibnamefont
  {Iqbal}}, \bibinfo {author} {\bibfnamefont {R.}~\bibnamefont {Thomale}},
  \bibinfo {author} {\bibfnamefont {F.}~\bibnamefont {{Parisen Toldin}}},
  \bibinfo {author} {\bibfnamefont {S.}~\bibnamefont {Rachel}}, \ and\ \bibinfo
  {author} {\bibfnamefont {J.}~\bibnamefont {Reuther}},\ }\bibfield  {title}
  {{\color{Gray}\small \bibinfo {title} {{Functional renormalization group for
  three-dimensional quantum magnetism}},\ }}\href {\doibase
  10.1103/PhysRevB.94.140408} {\bibfield  {journal} {\bibinfo  {journal} {Phys.
  Rev. B}\ }\textbf {\bibinfo {volume} {94}},\ \bibinfo {pages} {140408(R)}
  (\bibinfo {year} {2016})}\BibitemShut {NoStop}%
\bibitem [{\citenamefont {Popov}\ and\ \citenamefont
  {Fedotov}(1988)}]{Popov1988}%
  \BibitemOpen
  \bibfield  {author} {\bibinfo {author} {\bibfnamefont {V.~N.}\ \bibnamefont
  {Popov}}\ and\ \bibinfo {author} {\bibfnamefont {S.~A.}\ \bibnamefont
  {Fedotov}},\ }\bibfield  {title} {{\color{Gray}\small \bibinfo {title} {{The
  functional-integration method and diagram technique for spin systems}},\
  }}\href@noop {} {\bibfield  {journal} {\bibinfo  {journal} {Journal of
  Experimental and Theoretical Physics}\ }\textbf {\bibinfo {volume} {67}},\
  \bibinfo {pages} {535} (\bibinfo {year} {1988})}\BibitemShut {NoStop}%
\bibitem [{\citenamefont {Roscher}\ \emph {et~al.}(2019)\citenamefont
  {Roscher}, \citenamefont {Gneist}, \citenamefont {Scherer}, \citenamefont
  {Trebst},\ and\ \citenamefont {Diehl}}]{Roscher2019}%
  \BibitemOpen
  \bibfield  {author} {\bibinfo {author} {\bibfnamefont {D.}~\bibnamefont
  {Roscher}}, \bibinfo {author} {\bibfnamefont {N.}~\bibnamefont {Gneist}},
  \bibinfo {author} {\bibfnamefont {M.~M.}\ \bibnamefont {Scherer}}, \bibinfo
  {author} {\bibfnamefont {S.}~\bibnamefont {Trebst}}, \ and\ \bibinfo {author}
  {\bibfnamefont {S.}~\bibnamefont {Diehl}},\ }\bibfield  {title}
  {{\color{Gray}\small \bibinfo {title} {{Cluster functional renormalization
  group and absence of a bilinear spin liquid in the $J_1$-$J_2$ Heisenberg
  model}},\ }}\href {\doibase 10.1103/PhysRevB.100.125130} {\bibfield
  {journal} {\bibinfo  {journal} {Physical Review B}\ }\textbf {\bibinfo
  {volume} {100}},\ \bibinfo {pages} {125130} (\bibinfo {year}
  {2019})}\BibitemShut {NoStop}%
\bibitem [{\citenamefont {Katanin}(2004)}]{Katanin2004}%
  \BibitemOpen
  \bibfield  {author} {\bibinfo {author} {\bibfnamefont {A.~A.}\ \bibnamefont
  {Katanin}},\ }\bibfield  {title} {{\color{Gray}\small \bibinfo {title}
  {{Fulfillment of Ward identities in the functional renormalization group
  approach}},\ }}\href {\doibase 10.1103/PhysRevB.70.115109} {\bibfield
  {journal} {\bibinfo  {journal} {Phys. Rev. B}\ }\textbf {\bibinfo {volume}
  {70}},\ \bibinfo {pages} {115109} (\bibinfo {year} {2004})}\BibitemShut
  {NoStop}%
\bibitem [{\citenamefont {Keles}\ and\ \citenamefont {Zhao}(2018)}]{Keles2018}%
  \BibitemOpen
  \bibfield  {author} {\bibinfo {author} {\bibfnamefont {A.}~\bibnamefont
  {Keles}}\ and\ \bibinfo {author} {\bibfnamefont {E.}~\bibnamefont {Zhao}},\
  }\bibfield  {title} {{\color{Gray}\small \bibinfo {title} {{Absence of
  long-range order in a triangular spin system with dipolar interactions}},\
  }}\href {\doibase 10.1103/PhysRevLett.120.187202} {\bibfield  {journal}
  {\bibinfo  {journal} {Physical Review Letters}\ }\textbf {\bibinfo {volume}
  {120}},\ \bibinfo {pages} {187202} (\bibinfo {year} {2018})}\BibitemShut
  {NoStop}%
\bibitem [{\citenamefont {Kiese}\ \emph {et~al.}(2019)\citenamefont {Kiese},
  \citenamefont {Buessen}, \citenamefont {Hickey}, \citenamefont {Trebst},\
  and\ \citenamefont {Scherer}}]{Kiese2019}%
  \BibitemOpen
  \bibfield  {author} {\bibinfo {author} {\bibfnamefont {D.}~\bibnamefont
  {Kiese}}, \bibinfo {author} {\bibfnamefont {F.~L.}\ \bibnamefont {Buessen}},
  \bibinfo {author} {\bibfnamefont {C.}~\bibnamefont {Hickey}}, \bibinfo
  {author} {\bibfnamefont {S.}~\bibnamefont {Trebst}}, \ and\ \bibinfo {author}
  {\bibfnamefont {M.~M.}\ \bibnamefont {Scherer}},\ }\bibfield  {title}
  {{\color{Gray}\small \bibinfo {title} {{Emergence and stability of
  spin-valley entangled quantum liquids in moir\'e heterostructures}},\
  }}\href@noop {} {\  (\bibinfo {year} {2019})}\BibitemShut {NoStop}%
\bibitem [{\citenamefont {Affleck}\ \emph {et~al.}(1988)\citenamefont
  {Affleck}, \citenamefont {Zou}, \citenamefont {Hsu},\ and\ \citenamefont
  {Anderson}}]{Affleck1988}%
  \BibitemOpen
  \bibfield  {author} {\bibinfo {author} {\bibfnamefont {I.}~\bibnamefont
  {Affleck}}, \bibinfo {author} {\bibfnamefont {Z.}~\bibnamefont {Zou}},
  \bibinfo {author} {\bibfnamefont {T.}~\bibnamefont {Hsu}}, \ and\ \bibinfo
  {author} {\bibfnamefont {P.~W.}\ \bibnamefont {Anderson}},\ }\bibfield
  {title} {{\color{Gray}\small \bibinfo {title} {{SU(2) gauge symmetry of the
  large- U limit of the Hubbard model}},\ }}\href {\doibase
  10.1103/PhysRevB.38.745} {\bibfield  {journal} {\bibinfo  {journal} {Phys.
  Rev. B}\ }\textbf {\bibinfo {volume} {38}},\ \bibinfo {pages} {745} (\bibinfo
  {year} {1988})}\BibitemShut {NoStop}%
\bibitem [{\citenamefont {Kopietz}\ \emph {et~al.}(2010)\citenamefont
  {Kopietz}, \citenamefont {Bartosch},\ and\ \citenamefont
  {Sch{\"{u}}tz}}]{Kopietz2010a}%
  \BibitemOpen
  \bibfield  {author} {\bibinfo {author} {\bibfnamefont {P.}~\bibnamefont
  {Kopietz}}, \bibinfo {author} {\bibfnamefont {L.}~\bibnamefont {Bartosch}}, \
  and\ \bibinfo {author} {\bibfnamefont {F.}~\bibnamefont {Sch{\"{u}}tz}},\
  }\bibfield  {title} {{\color{Gray}\small \bibinfo {title} {{Functional
  Methods}},\ }}in\ \href@noop {} {\emph {\bibinfo {booktitle} {Lecture Notes
  in Physics}}},\ Vol.\ \bibinfo {volume} {798}\ (\bibinfo  {publisher}
  {Springer, Berlin, Heidelberg},\ \bibinfo {year} {2010})\ p.\ \bibinfo
  {pages} {147}\BibitemShut {NoStop}%
\bibitem [{Note1()}]{Note1}%
  \BibitemOpen
  \bibinfo {note} {{For our pf-FRG calculations, it is convenient to rotate the
  lattice such that it lies in the (111) plane. Correspondingly, when we refer
  to $\delimiter "426830A S^z S^z \delimiter "526930B $ correlations, we
  compute them in (111) direction.}}\BibitemShut {Stop}%
\bibitem [{\citenamefont {Buessen}\ \emph
  {et~al.}(2018{\natexlab{b}})\citenamefont {Buessen}, \citenamefont {Hering},
  \citenamefont {Reuther},\ and\ \citenamefont {Trebst}}]{Buessen2018}%
  \BibitemOpen
  \bibfield  {author} {\bibinfo {author} {\bibfnamefont {F.~L.}\ \bibnamefont
  {Buessen}}, \bibinfo {author} {\bibfnamefont {M.}~\bibnamefont {Hering}},
  \bibinfo {author} {\bibfnamefont {J.}~\bibnamefont {Reuther}}, \ and\
  \bibinfo {author} {\bibfnamefont {S.}~\bibnamefont {Trebst}},\ }\bibfield
  {title} {{\color{Gray}\small \bibinfo {title} {{Quantum spin liquids in
  frustrated spin-1 diamond antiferromagnets}},\ }}\href {\doibase
  10.1103/PhysRevLett.120.057201} {\bibfield  {journal} {\bibinfo  {journal}
  {Phys. Rev. Lett.}\ }\textbf {\bibinfo {volume} {120}},\ \bibinfo {pages}
  {057201} (\bibinfo {year} {2018}{\natexlab{b}})}\BibitemShut {NoStop}%
\bibitem [{\citenamefont {Elhajal}\ \emph {et~al.}(2002)\citenamefont
  {Elhajal}, \citenamefont {Canals},\ and\ \citenamefont
  {Lacroix}}]{Elhajal2002}%
  \BibitemOpen
  \bibfield  {author} {\bibinfo {author} {\bibfnamefont {M.}~\bibnamefont
  {Elhajal}}, \bibinfo {author} {\bibfnamefont {B.}~\bibnamefont {Canals}}, \
  and\ \bibinfo {author} {\bibfnamefont {C.}~\bibnamefont {Lacroix}},\
  }\bibfield  {title} {{\color{Gray}\small \bibinfo {title} {{Symmetry breaking
  due to Dzyaloshinsky-Moriya interactions in the kagom{\'{e}} lattice}},\
  }}\href {\doibase 10.1103/PhysRevB.66.014422} {\bibfield  {journal} {\bibinfo
   {journal} {Phys. Rev. B}\ }\textbf {\bibinfo {volume} {66}},\ \bibinfo
  {pages} {014422} (\bibinfo {year} {2002})}\BibitemShut {NoStop}%
\bibitem [{\citenamefont {Han}\ \emph {et~al.}(2016)\citenamefont {Han},
  \citenamefont {Norman}, \citenamefont {Wen}, \citenamefont
  {Rodriguez-Rivera}, \citenamefont {Helton}, \citenamefont {Broholm},\ and\
  \citenamefont {Lee}}]{Han2016}%
  \BibitemOpen
  \bibfield  {author} {\bibinfo {author} {\bibfnamefont {T.-H.}\ \bibnamefont
  {Han}}, \bibinfo {author} {\bibfnamefont {M.~R.}\ \bibnamefont {Norman}},
  \bibinfo {author} {\bibfnamefont {J.-J.}\ \bibnamefont {Wen}}, \bibinfo
  {author} {\bibfnamefont {J.~A.}\ \bibnamefont {Rodriguez-Rivera}}, \bibinfo
  {author} {\bibfnamefont {J.~S.}\ \bibnamefont {Helton}}, \bibinfo {author}
  {\bibfnamefont {C.}~\bibnamefont {Broholm}}, \ and\ \bibinfo {author}
  {\bibfnamefont {Y.~S.}\ \bibnamefont {Lee}},\ }\bibfield  {title}
  {{\color{Gray}\small \bibinfo {title} {{Correlated impurities and intrinsic
  spin-liquid physics in the kagome material herbertsmithite}},\ }}\href
  {\doibase 10.1103/PhysRevB.94.060409} {\bibfield  {journal} {\bibinfo
  {journal} {Phys. Rev. B}\ }\textbf {\bibinfo {volume} {94}},\ \bibinfo
  {pages} {060409(R)} (\bibinfo {year} {2016})}\BibitemShut {NoStop}%
\bibitem [{\citenamefont {Reuther}(2011)}]{Reuther2011}%
  \BibitemOpen
  \bibfield  {author} {\bibinfo {author} {\bibfnamefont {J.}~\bibnamefont
  {Reuther}},\ }\emph {\bibinfo {title} {{Frustrated quantum Heisenberg
  antiferromagnets: functional renormalization-group approach in
  auxiliary-fermion representation}}},\ \href
  {https://publikationen.bibliothek.kit.edu/1000023236} {Ph.D. thesis},\
  \bibinfo  {school} {Karlsruhe Institute of Technology} (\bibinfo {year}
  {2011})\BibitemShut {NoStop}%
\end{thebibliography}%


\appendix
\onecolumngrid

\section{Symmetries of the pseudo-fermion Hamiltonian}
\label{sec:appendix:symmetry}


\subsubsection{Local U(1) symmetry}
In this appendix, we present additional details on the derivation of symmetry constraints for the vertex parametrization. 
The action of the local U(1) symmetry is defined as 
\begin{equation}
\label{eq:appendix:symmetry:U(1)}
g_{\varphi_i} \left( \begin{array}{c}
f^\dagger_{i\alpha} \\ 
f^\nodagger_{i\alpha}
\end{array} \right) g_{\varphi_i}^{-1} = \left( \begin{array}{c}
e^{i\varphi_{i}} f^\dagger_{i\alpha} \\ 
e^{-i\varphi_i} f^\nodagger_{i\alpha}
\end{array} \right) \eqs, 
\end{equation}
i.e.~the transformation simply acts by multiplying the pseudo-fermion operators on lattice site $i$ with a phase factor of $\varphi_i$. 
In the notation that we are using throughout this section, the index $i$ refers to a lattice site (we use the letter $i$ to denote lattice sites since it is a common notation in literature, but the latter is simultaneously also being used as a symbol for the imaginary unit -- in such cases of double use the meaning should be clear to the reader from the context), and the index $\alpha$ denotes spin.
Applying this transformation to an arbitrary pseudo-fermion Hamiltonian leaves the Hamiltonian invariant. This is how it should be, since we are explicitly discussing a sub-group of the larger SU(2) gauge redundancy, whose existence we have confirmed earlier in Eq.~\eqref{eq:symmetries:pseudo-fermions:1} of the main article. 
In the current representation of spin operators, the invariance is easily seen from the fact that each single spin operator transforms as 
\begin{equation}
g_{\varphi_i} S_i^\mu g_{\varphi_i}^{-1} = g_{\varphi_i} \left( \frac{1}{2}f^\dagger_{i\alpha} \sigma_{\alpha\beta}^\mu f^\nodagger_{i\beta} \right) g_{\varphi_i}^{-1} 
= e^{i(\varphi_i-\varphi_i)} \left( \frac{1}{2}f^\dagger_{i\alpha} \sigma_{\alpha\beta}^\mu f^\nodagger_{i\beta} \right) = S_i^\mu  \eqs.
\end{equation}
Now, we examine the effect of the transformation on a two-point correlator. Consider to this end the correlation function (note that throughout the entire chapter we are only going to consider correlators of an equal number of creation and annihilation operators, see below)
\begin{equation}
\left< f^\dagger_{i'\tau'\alpha'} f^\nodagger_{i\tau\alpha} \right> =: G(i'\tau'\alpha' ; i\tau\alpha) \eqs.
\end{equation}
Two remarks are in order at this point. First of all, we have introduced an imaginary-time dependence, denoted by the additional index $\tau$. 
Moreover, our labeling suggests that we are dealing with the Green's function of the system. Yet, we slightly deviate from the conventional notion of a Green's function which would include time ordering on the imaginary time axis. 
We suppress time ordering, since we are ultimately interested in the implementation of symmetries in functional integral constructions, where time ordering becomes trivial after replacing fermionic operators with Grassmann numbers.
We may now apply the symmetry transformation to the correlator, which yields
\begin{equation}
\left< g_{\varphi_{i'}} g_{\varphi_i} f^\dagger_{i'\tau'\alpha'} f^\nodagger_{i\tau\alpha} g_{\varphi_i}^{-1} g_{\varphi_{i'}}^{-1} \right> 
= e^{i(\varphi_{i'}-\varphi_i)} \left< f^\dagger_{i'\tau'\alpha'} f^\nodagger_{i\tau\alpha} \right> \eqs, 
\end{equation}
which, upon Fourier transformation to Matsubara frequency space, becomes 
\begin{equation}
\left< g_{\varphi_{i'}} g_{\varphi_i} f^\dagger_{i'\omega'\alpha'} f^\nodagger_{i\omega\alpha} g_{\varphi_i}^{-1} g_{\varphi_{i'}}^{-1} \right> 
= e^{i(\varphi_{i'}-\varphi_i)} \left< f^\dagger_{i'\omega'\alpha'} f^\nodagger_{i\omega\alpha} \right> \eqs. 
\end{equation}
Since the transformation is a symmetry of the Hamiltonian, the correlator must be invariant under the transformation. 
For this to hold for arbitrary phase factors, the correlator has to be zero for $i' \neq i$. 
In a similar way, we may now investigate the four-point correlators. To keep the notation simple, we introduce composite indices, $n:=(i_n,\omega_n,\alpha_n)$, that simultaneously represent lattice site, Matsubara frequency, and spin index. We shall use this notation whenever suitable, but we may also fall back to explicitly stating all three indices separately when necessary. 
We define the four-point correlator as 
\begin{equation}
\left< f^\dagger_{1'} f^\dagger_{2'} f^\nodagger_1 f^\nodagger_2 \right> =: G(1', 2'; 1, 2) \eqs.
\end{equation}
In analogy to the two-point correlators, their transformation behavior under local U(1) is given by
\begin{equation}
\left< g_{\varphi_{i_{1'}}}g_{\varphi_{i_{2'}}}g_{\varphi_{i_{1}}}g_{\varphi_{i_{2}}} f^\dagger_{1'} f^\dagger_{2'} f^\nodagger_1 f^\nodagger_2 g_{\varphi_{i_{2}}}^{-1}g_{\varphi_{i_{1}}}^{-1}g_{\varphi_{i_{2'}}}^{-1}g_{\varphi_{i_{1'}}}^{-1} \right>  
= e^{i(\varphi_{i_{1'}}+\varphi_{i_{2'}}-\varphi_{i_1}-\varphi_{i_2})} \left< f^\dagger_{1'} f^\dagger_{2'} f^\nodagger_1 f^\nodagger_2 \right> \eqs. 
\end{equation}
In order for the phase factor to vanish, we have to impose bi-locality, meaning that the two incoming lattice sites (by incoming indices we mean those of annihilation operators) have to match the two outgoing lattice sites (by outgoing we refer to properties of creation operators). 
This leaves two non-zero combinations of lattice site indices: Either we pair up sites $i_{1'}$ and $i_1$ as well as $i_{2'}$ and $i_2$, or we match $i_{2'}$ and $i_1$ as well as $i_{1'}$ and $i_2$.


\subsubsection{Local particle-hole symmetry}
The action of the local particle-hole symmetry is given by 
\begin{equation}
\label{eq:appendix:symmetry:PH}
g_i \left( \begin{array}{c}
f^\dagger_{i\alpha} \\ 
f^\nodagger_{i\alpha}
\end{array} \right) g_i^{-1} = \left( \begin{array}{c}
\alpha f^\nodagger_{i\bar{\alpha}} \\ 
\alpha f^\dagger_{i\bar{\alpha}}
\end{array} \right) \eqs.
\end{equation}
The transformation locally exchanges creation and annihilation operators and reverses the spin. Note that unlike a physical particle-hole symmetry, however, this transformation is not anti-unitary. 
This symmetry is a subset of the SU(2) gauge redundancy and therefore holds for every pseudo-fermion Hamiltonian. 
When considering the action of the symmetry transformation on correlators, we note that the transformation acts trivially on the imaginary time index, as can be seen in the Heisenberg picture of operators. 
For a two-point correlator, the symmetry operation implies
\begin{equation}
\left< g_{i'}g_i f^\dagger_{i'\tau'\alpha'} f^\nodagger_{i\tau\alpha} g_i^{-1}g_{i'}^{-1} \right> = - \alpha'\alpha \left< f^\dagger_{i\tau\bar{\alpha}} f^\nodagger_{i'\tau'\bar{\alpha}'} \right> \eqs,
\end{equation}
where we assumed that the transformation is applied to both lattice sites involved. 
Fourier transformation yields the counterpart in frequency space, 
\begin{equation}
\left< g_{i'}g_i f^\dagger_{i'\omega'\alpha'} f^\nodagger_{i\omega\alpha} g_i^{-1}g_{i'}^{-1} \right>  = - \alpha'\alpha \left< f^\dagger_{i-\omega\bar{\alpha}} f^\nodagger_{i'-\omega'\bar{\alpha}'} \right> \eqs. 
\end{equation}
We now discuss the symmetry of bi-local four-point correlators (as a consequence of local U(1) symmetry, we have already seen that the correlator can only be non-zero when lattice sites are matched pairwise). Here, the local particle-hole transformation can be applied separately to either one of the pairs of lattice sites. Applying the transformation to lattice site $i_1$, we obtain
\begin{equation}
\left< g_{i_1} f^\dagger_{i_{1}\tau_{1'}\alpha_{1'}} f^\dagger_{i_{2}\tau_{2'}\alpha_{2'}} f^\nodagger_{i_{1}\tau_{1}\alpha_{1}} f^\nodagger_{i_{2}\tau_{2}\alpha_{2}} g_{i_1}^{-1} \right> = -\alpha_{1'}\alpha_{1} \left< f^\dagger_{i_{1}\tau_{1}\bar{\alpha}_{1}} f^\dagger_{i_{2}\tau_{2'}\alpha_{2'}} f^\nodagger_{i_{1}\tau_{1'}\bar{\alpha}_{1'}} f^\nodagger_{i_{2}\tau_{2}\alpha_{2}} \right> \eqs,
\end{equation}
which, upon Fourier transformation, becomes 
\begin{equation}
\left< g_{i_1} f^\dagger_{i_{1}\omega_{1'}\alpha_{1'}} f^\dagger_{i_{2}\omega_{2'}\alpha_{2'}} f^\nodagger_{i_{1}\omega_{1}\alpha_{1}} f^\nodagger_{i_{2}\omega_{2}\alpha_{2}} g_{i_1}^{-1} \right> 
= -\alpha_{1'}\alpha_{1} \left< f^\dagger_{i_{1}-\omega_{1}\bar{\alpha}_{1}} f^\dagger_{i_{2}\omega_{2'}\alpha_{2'}} f^\nodagger_{i_{1}-\omega_{1'}\bar{\alpha}_{1'}} f^\nodagger_{i_{2}\omega_{2}\alpha_{2}} \right> \eqs.
\end{equation}
Analogously, a second relation can be obtained from applying the symmetry relation to lattice site $i_2$, 
\begin{equation}
\left< g_{i_2} f^\dagger_{i_{1}\omega_{1'}\alpha_{1'}} f^\dagger_{i_{2}\omega_{2'}\alpha_{2'}} f^\nodagger_{i_{1}\omega_{1}\alpha_{1}} f^\nodagger_{i_{2}\omega_{2}\alpha_{2}} g_{i_2}^{-1} \right> 
= -\alpha_{2'}\alpha_{2} \left< f^\dagger_{i_{1}\omega_{1'}\alpha_{1'}} f^\dagger_{i_{2}-\omega_{2}\bar{\alpha}_{2}} f^\nodagger_{i_{1}\omega_{1}\alpha_{1}} f^\nodagger_{i_{2}-\omega_{2'}\bar{\alpha}_{2'}} \right> \eqs.
\end{equation}


\subsubsection{Lattice symmetries}
Lattice symmetries only act on the lattice site index. On the space of pseudo-fermions, they can be implemented as
\begin{equation}
\label{eq:appendix:symmetry:Lattice}
g_{T} \left( \begin{array}{c}
f^\dagger_{i\alpha} \\ 
f^\nodagger_{i\alpha}
\end{array} \right)g_{T}^{-1} = \left( \begin{array}{c}
f^\dagger_{T(i)\alpha} \\ 
f^\nodagger_{T(i)\alpha}
\end{array} \right) \eqs, 
\end{equation}
where $T$ is a lattice automorphism, which maps the lattice onto itself. 
Applied to a two-point correlation function, it acts as 
\begin{equation}
\left< g_T f^\dagger_{i'\tau'\alpha'} f^\nodagger_{i\tau\alpha} g^{-1}_T \right>
= \left< f^\dagger_{T(i')\tau'\alpha'} f^\nodagger_{T(i)\tau\alpha} \right> \eqs,
\end{equation}
which, upon Fourier transformation, becomes
\begin{equation}
\left< g_T f^\dagger_{i'\omega'\alpha'} f^\nodagger_{i\omega\alpha} g^{-1}_T \right>
= \left< f^\dagger_{T(i')\omega'\alpha'} f^\nodagger_{T(i)\omega\alpha} \right> \eqs. 
\end{equation}
On the bi-local four-point correlator, it acts as
\begin{align}
\left< g_T f^\dagger_{i_{1}\tau_{1'}\alpha_{1'}} f^\dagger_{i_{2}\tau_{2'}\alpha_{2'}} f^\nodagger_{i_{1}\tau_{1}\alpha_{1}} f^\nodagger_{i_{2}\tau_{2}\alpha_{2}} g^{-1}_T \right>
= \left< f^\dagger_{T(i_{1})\tau_{1'}\alpha_{1'}} f^\dagger_{T(i_{2})\tau_{2'}\alpha_{2'}} f^\nodagger_{T(i_{1})\tau_{1}\alpha_{1}} f^\nodagger_{T(i_{2})\tau_{2}\alpha_{2}} \right> \eqs, 
\end{align}
which is equivalent to the frequency dependent expression
\begin{align}
\left< g_T f^\dagger_{i_{1}\omega_{1'}\alpha_{1'}} f^\dagger_{i_{2}\omega_{2'}\alpha_{2'}} f^\nodagger_{i_{1}\omega_{1}\alpha_{1}} f^\nodagger_{i_{2}\omega_{2}\alpha_{2}} g^{-1}_T \right>
= \left< f^\dagger_{T(i_{1})\omega_{1'}\alpha_{1'}} f^\dagger_{T(i_{2})\omega_{2'}\alpha_{2'}} f^\nodagger_{T(i_{1})\omega_{1}\alpha_{1}} f^\nodagger_{T(i_{2})\omega_{2}\alpha_{2}} \right> \eqs.
\end{align}


\subsubsection{Time-reversal symmetry}
On pseudo-fermion operator level, time-reversal symmetry is implemented by the anti-unitary mapping 
\begin{equation}
\label{eq:appendix:symmetry:T}
g \left( \begin{array}{c}
f^\dagger_{i\alpha} \\ 
f^\nodagger_{i\alpha}
\end{array} \right) g^{-1} = \left( \begin{array}{c}
e^{i\pi \alpha/2} f^\dagger_{i\bar{\alpha}} \\ 
e^{-i\pi \alpha/2} f^\nodagger_{i\bar{\alpha}}
\end{array} \right) \eqs, 
\end{equation}
where the anti-linearity ensures that for all spin operators $g S^\mu g^{-1} = -S^\mu$. 
The transformation acts on the two-point correlator as 
\begin{equation}
\left< g f^\dagger_{i'\tau'\alpha'} f^\nodagger_{i\tau\alpha} g^{-1} \right>^* 
= e^{i\pi (\alpha-\alpha')/2} \left< f^\dagger_{i'\tau'\bar{\alpha}'} f^\nodagger_{i\tau\bar{\alpha}} \right>^* 
= \alpha'\alpha \left< f^\dagger_{i'\tau'\bar{\alpha}'} f^\nodagger_{i\tau\bar{\alpha}} \right>^* \eqs,
\end{equation}
where the star denotes complex conjugation and we have used that for a Hamiltonian, which is invariant under the anti-unitary transformation $g$, the thermal expectation value of an operator $A$ transforms as $\left< A \right> \rightarrow \left< g A g^{-1} \right>^*$. 
Furthermore, we have re-written the resulting phase factors as $e^{i\pi (\alpha-\alpha')/2}=\alpha'\alpha$. 
Fourier transformation of the expression yields
\begin{equation}
\label{eq:appendix:symmetry:T:2point}
\left< g f^\dagger_{i'\omega'\alpha'} f^\nodagger_{i\omega\alpha} g^{-1} \right>^* = 
\alpha'\alpha \left< f^\dagger_{i'-\omega'\bar{\alpha}'} f^\nodagger_{i-\omega\bar{\alpha}} \right>^* \eqs. 
\end{equation}
This symmetry is particularly helpful, because it links the real part and the imaginary part of the correlator (when comparing to the non-transformed expression). 
Although generically the correlator may be a complex number, time-reversal symmetry thus opens up the possibility to parametrize the two-point correlator by a real number instead of a complex number (or a pair of real numbers). 
Now we turn to the transformation behavior of the four-point correlator. The symmetry transformation of a bi-local four-point correlator is given by
\begin{align}
\left< g f^\dagger_{i_{1}\tau_{1'}\alpha_{1'}} f^\dagger_{i_{2}\tau_{2'}\alpha_{2'}} f^\nodagger_{i_{1}\tau_{1}\alpha_{1}} f^\nodagger_{i_{2}\tau_{2}\alpha_{2}} g^{-1} \right>^* 
&= e^{i\pi(\alpha_{1}+\alpha_{2}-\alpha_{1'}-\alpha_{2'})/2} \left< f^\dagger_{i_{1}\tau_{1'}\bar{\alpha}_{1'}} f^\dagger_{i_{2}\tau_{2'}\bar{\alpha}_{2'}} f^\nodagger_{i_{1}\tau_{1}\bar{\alpha}_{1}} f^\nodagger_{i_{2}\tau_{2}\bar{\alpha}_{2}} \right>^* \nonumber\\
&= \alpha_{1'}\alpha_{2'}\alpha_{1}\alpha_{2} \left< f^\dagger_{i_{1}\tau_{1'}\bar{\alpha}_{1'}} f^\dagger_{i_{2}\tau_{2'}\bar{\alpha}_{2'}} f^\nodagger_{i_{1}\tau_{1}\bar{\alpha}_{1}} f^\nodagger_{i_{2}\tau_{2}\bar{\alpha}_{2}} \right>^* \eqs.
\end{align}
In Matsubara frequency space the relation reads 
\begin{equation}
\label{eq:appendix:symmetry:T:4point}
\left< g f^\dagger_{i_{1}\omega_{1'}\alpha_{1'}} f^\dagger_{i_{2}\omega_{2'}\alpha_{2'}} f^\nodagger_{i_{1}\omega_{1}\alpha_{1}} f^\nodagger_{i_{2}\omega_{2}\alpha_{2}} g^{-1} \right>^* 
= \alpha_{1'}\alpha_{2'}\alpha_{1}\alpha_{2} \left< f^\dagger_{i_{1}-\omega_{1'}\bar{\alpha}_{1'}} f^\dagger_{i_{2}-\omega_{2'}\bar{\alpha}_{2'}} f^\nodagger_{i_{1}-\omega_{1}\bar{\alpha}_{1}} f^\nodagger_{i_{2}-\omega_{2}\bar{\alpha}_{2}} \right>^* \eqs.
\end{equation}
Just like for the two-point correlator, the symmetry can be used to make a connection between the real and the imaginary part of the correlator.


\subsubsection{Hermitian symmetry}
We assume that the Hamiltonian is Hermitian, i.e. invariant under a complex transposition. 
Since the correlation functions are scalar numbers, complex transposition is equivalent to complex conjugation, and we obtain for the two-point correlator
\begin{equation}
\left< f^\dagger_{i'\omega'\alpha'} f^\nodagger_{i\omega\alpha} \right>^* 
= \left< f^\dagger_{i'\omega'\alpha'} f^\nodagger_{i\omega\alpha} \right>^\dagger
= \left< f^\dagger_{i-\omega\alpha} f^\nodagger_{i'-\omega'\alpha'} \right> \eqs.
\end{equation}
For the four-point correlator, complex transposition yields
\begin{equation}
\left< f^\dagger_{i_{1}\omega_{1'}\alpha_{1'}} f^\dagger_{i_{2}\omega_{2'}\alpha_{2'}} f^\nodagger_{i_{1}\omega_{1}\alpha_{1}} f^\nodagger_{i_{2}\omega_{2}\alpha_{2}} \right>^*
= \left< f^\dagger_{i_{1}-\omega_{1}\alpha_{1}} f^\dagger_{i_{2}-\omega_{2}\alpha_{2}} f^\nodagger_{i_{1}-\omega_{1'}\alpha_{1'}} f^\nodagger_{i_{2}-\omega_{2'}\alpha_{2'}} \right> \eqs.
\end{equation}


\subsubsection{Green's functions}
In the previous subsections, we have discussed the transformation behavior of two-point and four-point correlation functions. 
In the main text of the article, we derive an efficient parametrization of the effective action, based on combinations of these symmetries. 
For better readability, we therefore collect the individual symmetry relations. 
We also add to the list (Matsubara) frequency conservation as a consequence of translation invariance in imaginary time. 
Furthermore, we include the trivial symmetry relation obtained from simultaneous exchange of both ingoing and outgoing particles. 
The symmetries for the two-point correlation function are given by 
\begin{align}
\tag{Local U(1)}
G(i'\omega'\alpha' ; i\omega\alpha) &= G(i\omega'\alpha' ; i\omega\alpha)\delta_{i'i} \\
\tag{Local PH}
G(i'\omega'\alpha';i\omega\alpha) &= -\alpha'\alpha G(i -\omega \bar{\alpha} ; i' -\omega' \bar{\alpha}') \\
\tag{Lattice}
G(i'\omega'\alpha' ; i\omega\alpha) &= G(T(i') \omega'\alpha' ; T(i) \omega\alpha) \\
\tag{Time reversal}
G(i'\omega'\alpha';i\omega\alpha) &= \alpha'\alpha G(i' -\omega' \bar{\alpha}' ; i -\omega \bar{\alpha})^* \\
\tag{Hermiticity}
G(i'\omega'\alpha';i\omega\alpha) &= G(i -\omega \alpha ; i' -\omega' \alpha')^* \\
\tag{Energy conservation}
G(i'\omega'\alpha' ; i\omega\alpha) &= G(i\omega'\alpha' ; i\omega\alpha)\delta_{\omega'\omega} \eqs.
\end{align}
The list of symmetries for the four-point correlator comprises
\begin{align}
\tag{Local U(1)}
G(1', 2'; 1, 2) &= G(1', 2'; 1, 2) \delta_{i_{1'} i_{1}}\delta_{i_{2'} i_{2}} - G(2', 1'; 1, 2) \delta_{i_{2'} i_{1}}\delta_{i_{1'} i_{2}} \\
\tag{Local PH 1}
G(1', 2'; 1, 2)\delta_{i_{1'} i_{1}}\delta_{i_{2'} i_{2}} &= -\alpha_{1'}\alpha_{1} G(i_{1}-\omega_{1}\bar{\alpha}_{1},i_{2}\omega_{2'}\alpha_{2'};i_{1}-\omega_{1'}\bar{\alpha}_{1'},i_{2}\omega_{2}\alpha_{2}) \\
\tag{Local PH 2}
G(1', 2'; 1, 2)\delta_{i_{1'} i_{1}}\delta_{i_{2'} i_{2}} &= -\alpha_{2'}\alpha_{2} G(i_{1}\omega_{1'}\alpha_{1'},i_{2}-\omega_{2}\bar{\alpha}_{2};i_{1}\omega_{1}\alpha_{1},i_{2}-\omega_{2'}\bar{\alpha}_{2'}) \\
\tag{Lattice}
G(1', 2'; 1, 2)\delta_{i_{1'} i_{1}}\delta_{i_{2'} i_{2}} &= G(T(i_{1})\omega_{1'}\alpha_{1'}, T(i_{2})\omega_{2'}\alpha_{2'}; T(i_{1})\omega_{1}\alpha_{1}, T(i_{2})\omega_{2}\alpha_{2}) \\
\tag{Time reversal}
G(1', 2'; 1, 2)\delta_{i_{1'} i_{1}}\delta_{i_{2'} i_{2}} &= \alpha_{1'}\alpha_{2'}\alpha_{1}\alpha_{2} G(i_{1}-\omega_{1'}\bar{\alpha}_{1'},i_{2}-\omega_{2'}\bar{\alpha}_{2'};i_{1}-\omega_{1}\bar{\alpha}_{1},i_{2}-\omega_{2}\bar{\alpha}_{2})^* \\
\tag{Hermiticity}
G(1', 2'; 1, 2)\delta_{i_{1'} i_{1}}\delta_{i_{2'} i_{2}} &= G(i_{1}-\omega_{1}\alpha_{1},i_{2}-\omega_{2}\alpha_{2};i_{1}-\omega_{1'}\alpha_{1'},i_{2}-\omega_{2'}\alpha_{2'})^* \\
\tag{Energy conservation}
G(1', 2'; 1, 2)\delta_{i_{1'} i_{1}}\delta_{i_{2'} i_{2}} &= G(1', 2'; 1, 2) \delta_{i_{1'} i_{1}}\delta_{i_{2'} i_{2}} \delta_{w_{1'}+w_{2'}-w_{1}-w_{2}} \\
\tag{Particle exchange}
G(1', 2'; 1, 2)\delta_{i_{1'} i_{1}}\delta_{i_{2'} i_{2}} &= G(2', 1'; 2, 1)\delta_{i_{1'} i_{1}}\delta_{i_{2'} i_{2}} \eqs.
\end{align}


\section{Symmetries of flow equations}
\label{sec:appendix:induction}
We will now explicitly show that the symmetries of the vertices stated in Sec.~\ref{sec:symmetriesFlowEquation} are indeed preserved during the $\Lambda$ flow. The following proofs are performed by induction, i.e., we assume that all symmetries are satisfied for the vertices on right hand sides of the flow equations and then show that the derivative of the vertex on the left hand side fulfills the symmetries as well. The initial step of the induction proof amounts to confirming that the symmetries are also obeyed in the initial conditions. Since this is trivial to show, we will omit this step. Note that our proofs are performed simultaneously for the two-particle vertex and for the self-energy; hence, when we show the symmetries for the two-particle vertex we may already assume the symmetries for the self-energy and vice versa. 

For reasons of brevity and since it has already been discussed in sufficient detail in the main text, we will not prove the condition in the first two lines of Eq.~(\ref{eq:pffrg:parametrization4point:1}) again, according to which the two-particle vertex is either purely real or purely imaginary. Furthermore, we will not show the antisymmetry of the self-energy in the frequency argument and the fact that the self-energy is purely imaginary. Both properties follow straightforwardly from the flow equation for the self-energy. Therefore, it only remains to be shown that the self-energy is proportional to $\delta_{\alpha_{1'}\alpha_1}$ in spin space.

We start with the proofs for the two-particle vertex symmetries as given in Eq.~(\ref{eq:symmetriesFlowEquation}). Our proofs here are similar to those presented in Ref.~\onlinecite{Reuther2011}. 
To start with, the symmetry 
\begin{equation}
\label{sym1}
\Gamma^{\Lambda}(1',2';1,2) = \Gamma^{\Lambda}(2',1';2,1)\\
\end{equation}
is obvious, since it only amounts to exchanging the ingoing and outgoing particles.

We continue proving the property
\begin{equation}
\label{sym3}
\Gamma^{\Lambda}(1',2';1,2) = \Gamma^{\Lambda}(1,2;1',2')^*\;.
\end{equation}
The induction is performed by first writing down the flow equation for $\Gamma^{\Lambda}(1,2;1',2')^*$ and then manipulating the right hand side,
\begin{equation}
\begin{split}
&\ \frac{d}{d\Lambda}\Gamma^{\Lambda}(1,2;1',2')^*
= \frac{1}{2\pi} \sum_{3,4}\Big[\Gamma^{\Lambda}(1,2;3,4)^* \Gamma^{\Lambda}(3,4;1',2')^*
-\Gamma^{\Lambda}(1,4;1',3)^*\Gamma^{\Lambda}(3,2;4,2')^*-(3\leftrightarrow4)\\
&+\Gamma^{\Lambda}(2,4;1',3)^*\Gamma^{\Lambda}(3,1;4,2')^*+(3\leftrightarrow4) \Big]G^{\Lambda}(\omega_3)^*S^{\Lambda}(\omega_4)^* \\
&\stackrel{\text{(\RNum{1})}}{=} \frac{1}{2\pi} \sum_{3,4}\Big[\Gamma^{\Lambda}(1',2';3,4) \Gamma^{\Lambda}(3,4;1,2)
-\Gamma^{\Lambda}(1',3;1,4)\Gamma^{\Lambda}(4,2';3,2)-(3\leftrightarrow4)\\
&+\Gamma^{\Lambda}(4,2';3,1)\Gamma^{\Lambda}(1',3;2,4)+(3\leftrightarrow4) \Big]G^{\Lambda}(\omega_3)S^{\Lambda}(\omega_4) \\
&\stackrel{\text{(\RNum{2})}}{=} \frac{1}{2\pi} \sum_{3,4}\Big[\Gamma^{\Lambda}(1',2';3,4) \Gamma^{\Lambda}(3,4;1,2)
-\Gamma^{\Lambda}(1',4;1,3)\Gamma^{\Lambda}(3,2';4,2)-(3\leftrightarrow4)\\
&+\Gamma^{\Lambda}(2',4;1,3)\Gamma^{\Lambda}(3,1';4,2)+(3\leftrightarrow4) \Big]G^{\Lambda}(\omega_3)S^{\Lambda}(\omega_4)
= \frac{d}{d\Lambda}\Gamma^{\Lambda}(1',2';1,2)\;.
\end{split}
\end{equation}
In step (\RNum{1}) we used Eq.~(\ref{sym3}), exchanged the orders of the two-particle vertices and used the property that the propagators are purely imaginary. In (\RNum{2}) we used Eq.~(\ref{sym1}) in the last line.\\

Next, we prove 
\begin{equation}
\label{sym2}
\Gamma^{\Lambda}(1',2';1,2) = \Gamma^{\Lambda}(-1',-2';-1,-2) \eqs,
\end{equation}
where $-n = \{i_n, -\omega_n, \alpha_n \}$ indicates that the sign of the frequency argument is flipped. The induction proof is performed in the same way as before,
\begin{equation}
\begin{split}
&\ \frac{d}{d\Lambda}\Gamma^{\Lambda}(-1',-2';-1,-2)
= \frac{1}{2\pi} \sum_{3,4}\Big[\Gamma^{\Lambda}(-1',-2';3,4) \Gamma^{\Lambda}(3,4;-1,-2)
-\Gamma^{\Lambda}(-1',4;-1,3)\Gamma^{\Lambda}(3,-2';4,-2)-(3\leftrightarrow4)\\
&+\Gamma^{\Lambda}(-2',4;-1,3)\Gamma^{\Lambda}(3,-1';4,-2)+(3\leftrightarrow4) \Big]
G^{\Lambda}(\omega_{3})S^{\Lambda}(\omega_{4}) \\
&\stackrel{\text{(\RNum{1})}}{=} \frac{1}{2\pi} \sum_{3,4}\Big[\Gamma^{\Lambda}(1',2';-3,-4) \Gamma^{\Lambda}(-3,-4;1,2)
-\Gamma^{\Lambda}(1',-4;1,-3)\Gamma^{\Lambda}(-3,2';-4,2)-(3\leftrightarrow4)\\
&+\Gamma^{\Lambda}(2',-4;1,-3)\Gamma^{\Lambda}(-3,1';-4,2)+(3\leftrightarrow4) \Big]
G^{\Lambda}(\omega_{3})S^{\Lambda}(\omega_{4})
\stackrel{\text{(\RNum{2})}}{=} \frac{d}{d\Lambda}\Gamma^{\Lambda}(1',2';1,2)\;.
\end{split}
\end{equation}
In step (\RNum{1}), we used Eq.~(\ref{sym2}) and in (\RNum{2}), the frequencies $\omega_3$ and $\omega_4$ have been substituted by $-\omega_3$ and $-\omega_4$. We also used the fact that the propagators are odd in frequency.

The last symmetry
\begin{equation}
\label{sym4}
\Gamma_{i_1 i_2}^{\Lambda}(1',2';1,2) = -\alpha_{2'}\alpha_{2} \Gamma_{i_1 i_2}^{\Lambda}(1',\bar{2};1,\bar{2}')
\end{equation}
is formulated for the vertex $\Gamma_{i_1 i_2}^{\Lambda}(1',2';1,2)$ which is related to $\Gamma^{\Lambda}(1',2';1,2)$ via
\begin{equation}
\Gamma^{\Lambda}(1',2';1,2)=\Gamma_{i_1 i_2}^{\Lambda}(1',2';1,2)\delta_{i_{1'}i_1}\delta_{i_{2'}i_2}-\Gamma_{i_1 i_2}^{\Lambda}(2',1';1,2)\delta_{i_{2'}i_1}\delta_{i_{1'}i_2}\eqs.
\end{equation}
Note that here the arguments $1'$, $2'$, etc. only include the frequency and the spin index but not the site. Furthermore, we use the notation $\bar{n} =\{-\omega_n, \Bar{\alpha}_n\}$. The flow equation for $\Gamma_{i_1 i_2}^{\Lambda}(1',2';1,2)$ has the form
\begin{equation}
\begin{split}
& \frac{d}{d\Lambda}\Gamma_{i_1 i_2}^{\Lambda}(1',2';1,2)
= \frac{1}{2\pi} \sum_{3,4}\Big[ \Gamma_{i_1 i_2}^{\Lambda}(1',2';3\,,4\,) \Gamma_{i_1 i_2}^{\Lambda}(3\,,4\,;1\,,2\,) + (3 \leftrightarrow 4)
-\sum_{j}\Gamma_{i_1 j}^{\Lambda}(1',4;1,3) \Gamma_{j i_2}^{\Lambda}(3,2';4,2) - (3 \leftrightarrow 4)\\
&+\Gamma_{i_1 i_2}^{\Lambda}(1',4;1,3) \Gamma_{i_2 i_2}^{\Lambda}(3,2';2,4) + (3 \leftrightarrow 4)
+\Gamma_{i_1 i_1}^{\Lambda}(1',4;3,1) \Gamma_{i_1 i_2}^{\Lambda}(3,2';4,2) + (3 \leftrightarrow 4)\\
&+\Gamma_{i_2 i_1}^{\Lambda}(2',4;3,1) \Gamma_{i_2 i_1}^{\Lambda}(3,1';2,4) + (3 \leftrightarrow 4)\Big]G^{\Lambda}(\omega_3) S^{\Lambda}(\omega_4)\eqs.
\end{split}
\end{equation}
The following proof is most conveniently carried out by considering the different channels of the flow equation individually. We label the five channels on the right hand side by the letters a to e, according to their order in the flow equation. We start with the particle-particle and crossed particle-hole channel.
\begin{equation}
\begin{split}
&\ -\alpha_2' \alpha_{2}\frac{d}{d\Lambda}\Gamma_{i_1 i_2}^{\Lambda,a+e}(1',\bar{2};1,\bar{2}')\\
&= - \frac{1}{2\pi} \sum_{3,4} \alpha_{2'}\alpha_2 \Big[\Gamma_{i_1 i_2}^{\Lambda}(1',\bar{2};3,4) \Gamma_{i_1 i_2}^{\Lambda}(3,4;1,\bar{2}')+(3\leftrightarrow4)
+\Gamma_{i_2 i_1}^{\Lambda}(\bar{2},4;3,1)\Gamma_{i_2 i_1}^{\Lambda}(3,1';\bar{2}',4)+(3\leftrightarrow4) \Big]G^{\Lambda}(\omega_3)S^{\Lambda}(\omega_4)\\
&\stackrel{\text{(\RNum{1})}}{=} - \frac{1}{2\pi} \sum_{3,4} \alpha_{2'}\alpha_2 \Big[\Gamma_{i_1 i_2}^{\Lambda}(1',\bar{2};3,4) \Gamma_{i_1 i_2}^{\Lambda}(3,4;1,\bar{2}')+(3\leftrightarrow4)
+\Gamma_{i_1 i_2}^{\Lambda}(4,\bar{2};1,3)\Gamma_{i_1 i_2}^{\Lambda}(1',3;4,\bar{2}')+(3\leftrightarrow4) \Big]G^{\Lambda}(\omega_3)S^{\Lambda}(\omega_4) \\
&\stackrel{\text{(\RNum{2})}}{=} - \frac{1}{2\pi} \sum_{3,4} \alpha_{2'}\alpha_2 \Big[
\Gamma_{i_1 i_2}^{\Lambda}(1',\bar{4};3,2) \Gamma_{i_1 i_2}^{\Lambda}(3,2';1,\bar{4})(-\alpha_2) \alpha_{4}\alpha_4 (-\alpha_{2'})+(3\leftrightarrow4)\\
&+\Gamma_{i_1 i_2}^{\Lambda}(4,\bar{3};1,2)\Gamma_{i_1 i_2}^{\Lambda}(1',2';4,\bar{3})(-\alpha_2) \alpha_{3}\alpha_3 (-\alpha_{2'})
+(3\leftrightarrow4) \Big]G^{\Lambda}(\omega_3)S^{\Lambda}(\omega_4)\\
&\stackrel{\text{(\RNum{3})}}{=} \frac{1}{2\pi} \sum_{3,4} \Big[\Gamma_{i_1 i_2}^{\Lambda}(1',4;3,2) \Gamma_{i_1 i_2}^{\Lambda}(3,2';1,4)+(3\leftrightarrow4)
+\Gamma_{i_1 i_2}^{\Lambda}(4,3;1,2)\Gamma_{i_1 i_2}^{\Lambda}(1',2';4,3)+(3\leftrightarrow4) \Big]G^{\Lambda}(\omega_3)S^{\Lambda}(\omega_4) \\
&\stackrel{\text{(\RNum{4})}}{=} \frac{1}{2\pi} \sum_{3,4} \Big[\Gamma_{i_2 i_1}^{\Lambda}(4,1';2,3) \Gamma_{i_2 i_1}^{\Lambda}(2',3;4,1)+(3\leftrightarrow4)
+\Gamma_{i_1 i_2}^{\Lambda}(4,3;1,2)\Gamma_{i_1 i_2}^{\Lambda}(1',2';4,3)+(3\leftrightarrow4) \Big]G^{\Lambda}(\omega_3)S^{\Lambda}(\omega_4) \\
&= \frac{d}{d\Lambda}\Gamma_{i_1 i_2}^{\Lambda,a+e}(1',2';1,2) \eqs.
\end{split}
\end{equation}
In steps (\RNum{1}) and (\RNum{4}) we used Eq.~(\ref{sym1}) and in (\RNum{2}) we applied Eq.~(\ref{sym4}). The product of spins reduces to one. Step (\RNum{3}) transforms the arguments $\bar{3}$ and $\bar{4}$ to $3$ and $4$ making use of the antisymmetry of the propagators in the frequency argument. We continue with one of the particle-hole terms.
\begin{equation}
\begin{split}
& -\alpha_{2'}\alpha_2 \frac{d}{d\Lambda}\Gamma_{i_1 i_2}^{\Lambda,c}(1',\bar{2};1,\bar{2}')=-\frac{1}{2\pi} \sum_{3,4} \alpha_{2'}\alpha_2 \Big[
\Gamma_{i_1 i_2}^{\Lambda}(1',4;1,3) \Gamma_{i_2 i_2}^{\Lambda}(3,\bar{2};\bar{2}',4)+(3\leftrightarrow4)
\Big]G^{\Lambda}(\omega_3)S^{\Lambda}(\omega_4) \\
&\stackrel{\text{(\RNum{1})}}{=} -\frac{1}{2\pi} \sum_{3,4} \alpha_{2'}\alpha_2 \Big[\Gamma_{i_1 i_2}^{\Lambda}(1',\bar{3};1,\bar{4}) \Gamma_{i_2 i_2}^{\Lambda}(3,\bar{4};\bar{2}',2)
\alpha_{4}\alpha_3 (-\alpha_{2}) \alpha_{4}
+(3\leftrightarrow4)
\Big]G^{\Lambda}(\omega_3)S^{\Lambda}(\omega_4) \\
&\stackrel{\text{(\RNum{2})}}{=}-\frac{1}{2\pi} \sum_{3,4} \alpha_{2'}\alpha_2 \Big[\Gamma_{i_1 i_2}^{\Lambda}(1',\bar{3};1,\bar{4}) \Gamma_{i_2 i_2}^{\Lambda}(\bar{4},3;2,\bar{2}') \alpha_{3} (-\alpha_{2})
+(3\leftrightarrow4)
\Big]G^{\Lambda}(\omega_3)S^{\Lambda}(\omega_4) \\
&\stackrel{\text{(\RNum{3})}}{=} \frac{1}{2\pi} \sum_{3,4} \alpha_{2'}\alpha_2 \Big[\Gamma_{i_1 i_2}^{\Lambda}(1',\bar{3};1,\bar{4}) \Gamma_{i_2 i_2}^{\Lambda}(\bar{4},2';2,\bar{3})
\quad\ \alpha_{3}(-\alpha_{2'})\alpha_{3} (-\alpha_{2})+(3\leftrightarrow4)
\Big]G^{\Lambda}(\omega_3)S^{\Lambda}(\omega_4) \\
&= \frac{1}{2\pi} \sum_{3,4} \Big[\Gamma_{i_1 i_2}^{\Lambda}(1',3;1,4) \Gamma_{i_2 i_2}^{\Lambda}(4,2';2,3)+(3\leftrightarrow4)
\Big]
G^{\Lambda}(\omega_3)S^{\Lambda}(\omega_4)= \frac{d}{d\Lambda}\Gamma_{i_1 i_2}^{\Lambda,c}(1',2';1,2)\eqs.
\end{split}
\end{equation}
Here steps (\RNum{1}) and (\RNum{3}) exploit Eq.~(\ref{sym4}), whereas (\RNum{2}) uses Eq.~(\ref{sym1}). Finally, we treat the remaining two particle-hole channels.
\begin{equation}
\begin{split}
&-\alpha_{2'}\alpha_2 \frac{d}{d\Lambda}\Gamma_{i_1 i_2}^{\Lambda,b+d}(1',\bar{2};1,\bar{2}')\\
&=- \frac{1}{2\pi} \sum_{3,4} \alpha_{2'}\alpha_2 \Big[- \sum_{j}\Gamma_{i_1 j}^{\Lambda}(1',4;1,3) \Gamma_{j i_2}^{\Lambda}(3,\bar{2};4,\bar{2}')-(3\leftrightarrow4)
+\Gamma_{i_1 i_1}^{\Lambda}(1',4;3,1) \Gamma_{i_1 i_2}^{\Lambda}(3,\bar{2};4,\bar{2}')+(3\leftrightarrow4)
\Big]G^{\Lambda}(\omega_3)S^{\Lambda}(\omega_4) \\
&= \frac{1}{2\pi} \sum_{3,4} \alpha_{2'}\alpha_2 \Big[-\sum_{j}\Gamma_{i_1 j}^{\Lambda}(1',4;1,3) \Gamma_{j i_2}^{\Lambda}(3,2';4,2)(-\alpha_2) (-\alpha_{2'})-(3\leftrightarrow4)\\
&+\Gamma_{i_1 i_1}^{\Lambda}(1',4;3,1) \Gamma_{i_1 i_2}^{\Lambda}(3,2';4,2)(-\alpha_2) (-\alpha_{2'})
+(3\leftrightarrow4)
\Big]G^{\Lambda}(\omega_3)S^{\Lambda}(\omega_4) =\frac{d}{d\Lambda}\Gamma_{i_1 i_2}^{\Lambda,b+d}(1',2';1,2)\eqs.
\end{split}
\end{equation}
We have only used Eq.~(\ref{sym4}) in this proof. This concludes our proofs for the symmetry properties of the two-particle vertices.

We finally show that the self-energy is proportional to $\delta_{\alpha_{1'}\alpha_1}$ in spin space. Since it will be needed below, we first express the symmetries of the two-particle vertex using the parametrization of Eq.~(\ref{eq:pffrg:parametrization4point}),
\begin{align}
\Gamma^{\mu \nu}_{i_1 i_2}(\omega_{1'},\omega_{2'};\omega_{1},\omega_{2}) = \Gamma^{\nu \mu}_{i_2 i_1}(\omega_{2'},\omega_{1'};\omega_{2},\omega_{1})\;&,\quad
\Gamma^{\mu \nu}_{i_1 i_2}(\omega_{1'},\omega_{2'};\omega_{1},\omega_{2}) = \Gamma^{\mu \nu}_{i_1 i_2}(\omega_{1},\omega_{2};\omega_{1'},\omega_{2'})^*\;,\nonumber\\
\Gamma^{\mu \nu}_{i_1 i_2}(\omega_{1'},\omega_{2'};\omega_{1},\omega_{2}) = \Gamma^{\mu \nu}_{i_1 i_2}(-\omega_{1'},-\omega_{2'};-\omega_{1},-\omega_{2})\;&,\quad
\Gamma^{\mu \nu}_{i_1 i_2}(\omega_{1'},\omega_{2'};\omega_{1},\omega_{2}) = - \xi(\nu) \Gamma^{\mu \nu}_{i_1 i_2}(\omega_{1'},-\omega_{2};\omega_{1},-\omega_{2'})\;,
\end{align}
where the relations appear in the same order as in Eq.~(\ref{eq:symmetriesFlowEquation}). Using the fact that the two-particle vertices are either purely real or purely imaginary [see Eq.~(\ref{eq:pffrg:parametrization4point:1})] these conditions directly lead to
\begin{align}
\Gamma^{\mu\nu,\Lambda}_{i_1 i_2}(\omega_{1},\omega_{2};\omega_{1},\omega_{2}) = \Gamma^{\mu\nu,\Lambda}_{i_1 i_2}(\omega_{1},\omega_{2};\omega_{1},\omega_{2})^* 
\;\implies\; \Gamma^{\mu 0,\Lambda}_{i_1 i_2}(\omega_{1},\omega_{2};\omega_{1},\omega_{2}) = \Gamma^{0 \mu,\Lambda}_{i_1 i_2}(\omega_{1},\omega_{2};\omega_{1},\omega_{2}) =0\;\text{for $\mu=1,2,3$}
\label{mu00}
\end{align}
and
\begin{align}
&\Gamma^{\mu\nu,\Lambda}_{i_1 i_1}(\omega_{1},\omega_{2};\omega_{2},\omega_{1}) = \Gamma^{\mu\nu,\Lambda}_{i_1 i_1}(\omega_{2},\omega_{1};\omega_{1},\omega_{2})^{*} 
= \Gamma^{\nu\mu,\Lambda}_{i_1 i_1}(\omega_{1},\omega_{2};\omega_{2},\omega_{1})^{*}\nonumber\\
&\implies \Gamma^{\mu 0,\Lambda}_{i_1 i_1}(\omega_{1},\omega_{2};\omega_{2},\omega_{1}) = -\Gamma^{0 \mu,\Lambda}_{i_1 i_1}(\omega_{1},\omega_{2};\omega_{2},\omega_{1})
\;\text{for $\mu=1,2,3$}\nonumber\\
&\implies\Gamma^{\mu \nu,\Lambda}_{i_1 i_1}(\omega_{1},\omega_{2};\omega_{2},\omega_{1}) = \Gamma^{\nu \mu,\Lambda}_{i_1 i_1}(\omega_{1},\omega_{2};\omega_{2},\omega_{1})\;\text{for $\mu,\nu \in \{1,2,3\}$ or $\mu,\nu = 0$}\eqs.
\label{munu}
\end{align}
These relations will now be used to prove the diagonal form of the self-energy in its spin arguments as shown in Eq.~(\ref{eq:pffrg:parametrization2point}). The proof is based on the flow equation for the self-energy as given in Eq.~(\ref{eq:pffrg:flow:1particle}) where the two-particle vertex on the right hand side is rewritten using the parametrization of Eq.~(\ref{eq:pffrg:parametrization4point}). As before, our induction proof is performed by assuming that the symmetry to be shown is fulfilled for the vertex functions on the right hand side (i.e. that the self-energy is proportional to $\delta_{\alpha_{1'}\alpha_1}$ in our case) and then verifying that the symmetry also holds for the derivative on the left hand side. The proof reads
\begin{equation}
\begin{split}
&\frac{d}{d \Lambda} \Sigma^{\Lambda}(\omega_1,\alpha_{1'}, \alpha_{1})
=-\frac{1}{2\pi} \int d\omega_2 \sum_{\alpha_2}\Big[\sum_{i_2}\Gamma^{\mu\nu,\Lambda}_{i_1 i_2}(\omega_{1},\omega_{2};\omega_{1},\omega_{2}) \sigma_{\alpha_{1'}\alpha_{1}}^{\mu} \sigma_{\alpha_{2}\alpha_{2}}^{\nu} \\
&- \Big(\sum_{\substack{\mu > \nu\\ \mu,\nu \neq 0}}\Gamma^{\mu\nu,\Lambda}_{i_1 i_1}(\omega_{1},\omega_{2};\omega_{2},\omega_{1})(\sigma_{\alpha_{1'}\alpha_{2}}^{\mu} \sigma_{\alpha_{2}\alpha_{1}}^{\nu} +\sigma_{\alpha_{1'}\alpha_{2}}^{\nu} \sigma_{\alpha_{2}\alpha_{1}}^{\mu})
+ \sum_{\mu \neq 0} \Gamma^{\mu0,\Lambda}_{i_1 i_1}(\omega_{1},\omega_{2};\omega_{2},\omega_{1})(\sigma_{\alpha_{1'}\alpha_{2}}^{\mu} \sigma_{\alpha_{2}\alpha_{1}}^{0} -\sigma_{\alpha_{1'}\alpha_{2}}^{0} \sigma_{\alpha_{2}\alpha_{1}}^{\mu})\\
&+\sum_{\mu} \Gamma^{\mu\mu,\Lambda}_{i_1 i_1}(\omega_{1},\omega_{2};\omega_{2},\omega_{1}) \sigma_{\alpha_{1'}\alpha_{2}}^{\mu} \sigma_{\alpha_{2}\alpha_{1}}^{\mu}
\Big) \Big] S^{\Lambda}(\omega_2)\\
&=-\frac{1}{2\pi} \int d\omega_2 \Big[2\sum_{i_2}\Gamma^{00,\Lambda}_{i_1 i_2}(\omega_{1},\omega_{2};\omega_{1},\omega_{2}) 
-\sum_{\mu} \Gamma^{\mu\mu,\Lambda}_{i_1 i_1}(\omega_{1},\omega_{2};\omega_{2},\omega_{1})
\Big) \Big]\delta_{\alpha_{1'}\alpha_{1}} S^{\Lambda}(\omega_2)\eqs.
\end{split}
\end{equation}
The first equation is just the flow equation for the self-energy using the parametrization of Eq.~(\ref{eq:pffrg:parametrization4point}). The right hand side of this equation has four terms which will be discussed individually. [For convenience, the terms of the sum $\sum_{\mu\nu}$ have been split up in the second and third lines and the properties of Eq.~(\ref{munu}) have been used in the second and third terms.] The first term can be treated using Eq.~(\ref{mu00}) and the fact that the trace of Pauli matrices vanishes. One finds that only the $\Gamma^{00,\Lambda}$ term remains in the last line. The second term vanishes because $\{ \sigma^{\mu}, \sigma^{\nu} \} = 2 \delta_{\mu \nu} \sigma^{0}$ for $\mu,\nu \in \{1,2,3\}$. The third term vanishes too and for the fourth term we use $(\sigma^{\mu})^2 = \sigma^0$. We, hence, observe that only terms proportional to $\delta_{\alpha_{1'}\alpha_{1}}$ survive. This concludes our proof that the spin arguments of the self-energy satisfy the form given in Eq.~(\ref{eq:pffrg:parametrization2point}).

\end{document}